\documentclass[aps,prd,groupedaddress, tightenlines,preprint,eqsecnum,%
nofootinbib]{revtex4}
\usepackage{graphicx,epsf,amssymb,amsbsy,amsfonts,amssymb,amsmath}
\usepackage[usenames, dvipsnames]{color}

\newif\ifdraft
\drafttrue
\newif\ifpreprint
\preprinttrue

\def\tree{{\rm tree}}

\def\eps{\epsilon}
\def\e{\epsilon}

\def\nn{\nonumber}

\def\NeqFour{{{\cal N}=4}}

\def\NeqEight{{{\cal N}=8}}

\def\be{\begin{equation}}
\def\ee{\end{equation}}
\def\bea{\begin{eqnarray}}
\def\eea{\end{eqnarray}}
\def\ba{\begin{eqnarray}}
\def\ea{\end{eqnarray}}
\newcommand{\eq}[1]{eq.~(\ref{eq:#1})}
\def\eqn#1{eq.~(\ref{#1})}

\def\eqns#1#2{eqs.~(\ref{#1}) and~(\ref{#2})}

\def\fig#1{fig.~\ref{#1}}

\def\figs#1#2{Figs.~{\ref{#1}} and {\ref{#2}}}

\def\sect#1{section~{\ref{#1}}}
\def\Sect#1{Section~{\ref{#1}}}

\def \Tr {\mathop{\rm Tr}\nolimits}

\def\IC{\mathbb{C}}

\def\IZ{\mathbb{Z}}
\def\IR{\mathbb{R}}
\def\IP{\mathbb{P}}

\def\Vegas{{\sc Vegas}}

\begin{document}
\thispagestyle{empty}

\ifpreprint
\noindent
SLAC--PUB--15269 \hfill 
UCLA/12/TEP/104 $\null\hskip0.35cm\null$ \hfill 
SU-ITP-12/24 $\null\hskip0.35cm\null$\\
CERN-PH-TH/2012-288 \hfill 
\\
\fi

\begingroup\centering
{\Large\bfseries\mathversion{bold}
$D=5$ maximally supersymmetric\\
 Yang-Mills theory
diverges at six loops
\par}%
\vspace{7mm}

\begingroup\scshape\large
Zvi~Bern,$^{(1)}$ John~Joseph~Carrasco,$^{(2)}$ Lance~J.~Dixon,$^{(3)}$\\
Michael~R.~Douglas,$^{(4,5)}$
Matt von~Hippel$^{(4)}$\\
and Henrik Johansson$^{(6)}$\\
\endgroup
\vspace{8mm}
\begingroup\small
$^{(1)}$ \emph{Department of Physics and Astronomy, UCLA, 
Los Angeles, CA 90095-1547, USA} \\
\endgroup

\begingroup\small
$^{(2)}$ \emph{Stanford Institute for Theoretical Physics 
and Department of Physics, 
Stanford University, Stanford, CA 94305-4060, USA} \\
\endgroup

\begingroup\small
$^{(3)}$ \emph{SLAC National Accelerator Laboratory,
Stanford University, Stanford, CA 94309, USA} \\
\endgroup

\begingroup\small
$^{(4)}$ \emph{Simons Center for Geometry and Physics,
Stony Brook University,
Stony Brook NY 11794 }\\
\endgroup

\begingroup\small
$^{(5)}$ \emph{I.H.E.S, Bures-sur-Yvette, France 91440}\\
\endgroup

\begingroup\small
$^{(6)}$ \emph{Theory Division, Physics Department, CERN, CH-1211 Geneva 23, 
               Switzerland}\\
\endgroup

\vspace{0.6cm}
\begingroup\small
 {\tt bern@ucla.edu}\,,\;
 {\tt jjmc@stanford.edu}\,,\;
 {\tt lance@slac.stanford.edu}\,,\; 
 {\tt mdouglas@scgp.stonybrook.edu}\,,\; 
 {\tt matthew.vonhippel@stonybrook.edu}\,,\; 
 {\tt henrik.johansson@cern.ch}
\endgroup
\vspace{1.2cm}

\textbf{Abstract}\vspace{5mm}\par
\begin{minipage}{14.7cm}

The connection of maximally supersymmetric Yang-Mills theory to the $(2,0)$
theory in six dimensions has raised the possibility that it might be
perturbatively ultraviolet finite in five dimensions.  We test this hypothesis
by computing the coefficient of the first potential ultraviolet divergence
of planar (large $N_c$) maximally supersymmetric Yang-Mills theory in $D=5$, 
which occurs at six loops.  We show that the coefficient is nonvanishing.
Furthermore, the numerical value of the divergence
falls very close to an approximate exponential formula based on
the coefficients of the divergences through five loops.
This formula predicts the approximate values of the ultraviolet divergence
at loop orders $L > 6$ in the critical dimension $D=4+6/L$.  To obtain the
six-loop divergence we first construct the planar six-loop four-point
amplitude integrand using generalized unitarity.
The ultraviolet divergence follows from a set of vacuum integrals, which
are obtained by expanding the integrand in the external momenta.
The vacuum integrals are integrated via sector decomposition, using a 
modified version of the {\sc FIESTA} program.

\end{minipage}\par
\endgroup

\newpage

\tableofcontents

\section{Introduction}

Recent years have seen impressive progress in computing scattering
amplitudes in general gauge and gravity theories (see, for example, the
recent reviews~\cite{RecentAmplitudeReviews}).  The progress has been
especially great for maximally supersymmetric Yang-Mills theory (MSYM),
a theory with sixteen supercharges whose $D=4$ version is
$\NeqFour$ super-Yang-Mills theory.
One application has been to study the ultraviolet (UV)
properties of both gauge and gravity theories.
The all-loop ultraviolet finiteness of $\NeqFour$
super-Yang-Mills theory in $D=4$ was established in the
1980's~\cite{Mandelstam}.  In dimensions $D>4$, explicit
unitarity-based amplitude computations in MSYM in the 1990's~\cite{BRY,BDDPR}
showed that its degree of convergence was a bit better than had been
anticipated.  The results suggested that the correct finiteness bound
for MSYM in $D$ dimensions at $L$ loops is
\begin{equation}
D < 4 + \frac{6}{L} \hskip2cm (L \ge 2) \,.
\label{eq:FinitenessBound}
\end{equation}
This bound is consistent with all-loop ultraviolet finiteness in $D=4$.
The bound~(\ref{eq:FinitenessBound}) has been confirmed to all
loop orders~\cite{HoweStelleRevisited} using harmonic
superspace~\cite{HarmonicSuperspace}.  The case $L=1$ is an 
exception; at one loop the first divergence is in $D=8$, not
$D=10$~\cite{GSB}.  Interestingly, maximal $\NeqEight$ supergravity
follows precisely the same finiteness bound, at least through four
loops~\cite{PreviousGravityUV,CompactThree,CK4l}.

An important remaining question is whether the bound is saturated or
not; that is, whether the coefficient of the potential logarithmic
divergence in the critical dimension $D=4+6/L$ is nonzero or not, for
each loop order.  On the one hand, if the theory contains some unknown
or hidden symmetry, then the coefficient in $D=4+6/L$ could vanish,
leading to a higher critical dimension than expected at some loop
order. That is, at a loop order affected by the symmetry, the
lowest dimension with an ultraviolet divergence would be surprisingly
high.  On the other hand, if the bound is saturated, it proves that no
additional hidden symmetries exist that alter the degree of divergence
--- at least through the loop orders explored.

The only known reliable means for answering such a question is to explicitly 
evaluate the ultraviolet divergence of an appropriate on-shell multi-loop
scattering amplitude in the expected critical dimension.
Such a computation can be performed either within the large-$N_c$, or planar,
limit of the theory with gauge group $SU(N_c)$, or for a general gauge
group including all subleading terms in the $1/N_c$ expansion.
We know from these
computations~\cite{BDDPR, PreviousGravityUV,CompleteFourLoopYM,%
PlanarFive,FiveLoopNew}
that the bound~(\ref{eq:FinitenessBound})
is indeed saturated through at least five loops.  Interestingly,
certain subleading-in-$N_c$ terms (e.g.~double-trace terms),
do have an improved behavior at three loops and
beyond~\cite{CompleteFourLoopYM,SubleadingColorImprovement,CK4l}.
In this paper, we will only consider the leading-color (planar) terms.

If the bound~(\ref{eq:FinitenessBound}) is saturated, then
\eq{FinitenessBound} implies that in $D=5$ a divergence should
first appear at six loops. A primary purpose of this paper is to
compute the coefficient of this divergence.
The case of $D=5$ is especially interesting
because this theory has a UV completion, the $(2,0)$ theory in $D=6$.
The $(2,0)$ theory has no Lagrangian description; rather its
existence follows from arguments in string theory and M
theory~\cite{Witten1995zh}.  This connection suggests that UV
divergences in $D=5$ could give us information about the $(2,0)$
theory \cite{DouglasConjecture}.  Of course, a low-energy effective
theory usually has UV divergences, and by itself this connection does
not lead to constraints.  However, the present example is somewhat unique in
that the $(2,0)$ theory has neither a dimensionless coupling constant nor a
preferred scale, so that seemingly different $D=5$ quantities turn out
to be related nonperturbatively.  For example, the Kaluza-Klein
modes in $D=5$, arising from the compactification of the $(2,0)$ theory
on a circle, can also be identified with solitons in the gauge
theory~\cite{KKSolitons,DouglasConjecture,Lambert} (solutions
corresponding to instantons in $D=4$). 

In ref.~\cite{DouglasConjecture}
these aspects were discussed in the context of $S$-duality of the $D=4$
$\NeqFour$ super-Yang-Mills theory obtained by compactification of
$D=5$ MSYM on a circle, which can also be thought of as compactification
of the $(2,0)$ theory on a two-torus.
In this construction $S$-duality has a geometric origin as an exchange of
the two sides of the torus.  This argument can be re-expressed in $D=5$
terms, and UV divergences in $D=5$ can potentially violate the $D=4$
$S$-duality.  An alternative argument suggesting finiteness~\cite{Lambert}
is based on the soliton-Kaluza-Klein correspondence for
the compactified $(2,0)$ theory in the phase where the gauge
symmetry is broken by separating the multiple branes used in its
construction.
Although these arguments do not prove that there are no UV divergences,
they do motivate the question. In the present paper we definitively
answer the question, by computing the numerical coefficient of the
potential divergence in planar MSYM at six loops.

We find that the bound~(\ref{eq:FinitenessBound}) is indeed saturated
for $L=6$ and $D=5$; that is, the divergence has a nonzero coefficient.
Somewhat surprisingly, we also find that, through at least six loops,
the numerical values of the leading-color planar critical-dimension
divergences can be fit accurately to a simple exponential Ansatz.
Although we do not understand the origin of this simple functional
form, it does have useful consequences: It gives us additional confidence
that we have correctly computed the six-loop divergence and that it is
non-zero.  Moreover, by extrapolating to higher-loop orders it allows
us to predict the approximate numerical values of the divergences for
$L \ge 7$ in their critical dimensions.  This result suggests that even
outside of four dimensions, MSYM has a surprisingly simple
structure, reflected in the simple pattern of ultraviolet divergences.

Until recently, a direct evaluation of the six-loop ultraviolet
properties of MSYM would have been out of reach.
However, a combination of advances has made it possible.
Many breakthroughs in understanding the structure of integrands for
multi-loop amplitudes now allow for a rather straightforward construction
of the integrand for the six-loop four-point
amplitude in planar MSYM~\cite{RecentAmplitudeReviews}.
In addition to the integrand construction presented here, two
recent papers give the same integrands in $D=4$~\cite{BourjailyRep,EdenRep},
albeit presented somewhat differently.  Our construction is also valid
for loop momenta spanning the full five dimensions.
Well-developed techniques for extracting UV divergences from
amplitudes~\cite{Vladimirov,MarcusSagnotti,PreviousGravityUV,FiveLoopNew}
then allow us to express the six-loop divergence in terms of
a relatively small number of vacuum integrals.  Furthermore, a set of
integral consistency relations~\cite{FourLoopGravity}, related to
integration by parts identities~\cite{CTIBP}, allows us to
reduce the number of integrals further.  It also provides for nontrivial
cross-checks on numerical evaluations of the integrals and an 
independent means for estimating numerical integration uncertainty.

The final integrals obtained, after imposing the consistency relations,
are nevertheless quite challenging to evaluate analytically.
Instead, we make use of the major advances in
numerical integration techniques.  A long-standing challenge of
numerically evaluating Euclidean Feynman integrals has been addressed by
a computational technique called sector
decomposition~\cite{BinothHeinrichSector}, implemented in several
software packages~\cite{BognerWeinzierl,CarterSecDec, Fiesta}
including the {\sc FIESTA} package~\cite{Fiesta2}, used here.
To our knowledge, the sector decomposition technique has not previously been
applied at such a high loop order.  However, when applied to this problem it
leads successfully to integrals that can be evaluated on a
moderate size (1000 node) cluster in a few days.

The remainder of this paper is organized as follows.
In \sect{ConstructingIntegrandSection} we outline the derivation of the
integrand. \Sect{VacuumSection}
describes the procedure for extracting the UV divergence from the
amplitude in terms of a set of vacuum integrals.
\Sect{SectorDecompositionSection} explains the sector decomposition
method, and \sect{ComputationalDetailsSection} describes details of
the numerical evaluation.  In \sect{ConclusionsSection} we give our
conclusions, and comment on the feasibility of evaluating
the seven-loop integrals that might be required to check whether the
coefficient of the first potential counterterm of $\NeqEight$
supergravity vanishes.

\section{Constructing the integrand}
\label{ConstructingIntegrandSection}

Our study of the UV properties of the planar MSYM amplitude in $D=5$
begins by constructing the integrand of the six-loop four-point
amplitude.  Because we need the integrand in five dimensions, 
we must ensure that our construction is valid for loop momenta
inhabiting five spacetime dimensions.  (We can always take the external
momenta to be four-dimensional, and assign helicities to the external
gluons if desired.)  The five-dimensional validity of the integrand is
accomplished by verifying unitarity cuts in higher dimensions, which 
we have done on a large class of cuts. In
addition, to extract the UV divergences, we prefer a local form for
the integrand, in which the only denominator factors are standard
Feynman propagators. To find the desired representation, we use generalized
unitarity, a particularly effective
general purpose refinement of the unitarity method~\cite{UnitarityMethod}.
(For recent reviews of this method see refs.~\cite{UnitarityReviews}.) 
Our form for the integrand differs somewhat from recent ones based on
four-dimensional constructions~\cite{BourjailyRep,EdenRep}. However,
we have confirmed analytic agreement in any dimension with the
form in ref.~\cite{EdenRep} (which is also known to agree
with that in ref.~\cite{BourjailyRep}). 

We will focus on the leading-color, planar contribution to the six-loop
amplitude in $SU(N_c)$ gauge theory, which has the same color structure
as the tree amplitude, up to overall factors of the number of colors, $N_c$.
The color-decomposed form of the planar contribution to the 
$L$-loop four-point amplitude is,
\begin{eqnarray}
{\cal A}_4^{(L)} & = & g^{2}
 \Bigl[ { g^2 N_c } \Bigr]^{L}
 \sum_{\rho\in S_3}
\Tr( T^{a_{\rho(1)}} T^{a_{\rho(2)}}  T^{a_{\rho(3)}}  T^{a_{\rho(4)}} )
     A_4^{(L)}(\rho(1), \rho(2), \rho(3), \rho(4))\,, \hskip 2 cm
\label{LeadingColorDecomposition}
\end{eqnarray}
where $A_4^{(L)}$ is an $L$-loop color-ordered partial amplitude.
The sum runs over non-cyclic
permutations, $\rho$, of the external legs.  In this expression we
have suppressed momentum and helicity labels, leaving only the
indices identifying the external legs.  This decomposition holds for
any set of external particles from the full gauge supermultiplet.  

We will not describe our construction of the
six-loop amplitude in any detail; it is similar to the
construction of the five-loop planar amplitude given in
ref.~\cite{PlanarFive}.  Integrands of planar amplitudes in 
MSYM are relatively simple to obtain because dual conformal
symmetry severely restricts their
form~\cite{MagicIdentities,PlanarFour,KorchemskyZero,PlanarFive,NimaRecursion}.
Although dimensional regularization breaks dual conformal
invariance, it does so mildly at the level of the integrand.
Indeed, the integrands of loop
amplitudes in planar MSYM are known to have the same simple
properties under dual conformal transformation in dimensions $D \le 6$
(and likely in all dimensions $D\le 10$) as they have in four
dimensions~\cite{DualConfFiveD,SixDCheck,LoopConformal}.
The only breaking of dual conformal invariance comes from the integration
measure.  This allows us to use dual conformal symmetry to guide the
construction of the integrand, even outside of four dimensions.

\begin{figure}[th]
\centerline{\epsfxsize 5.5 truein \epsfbox{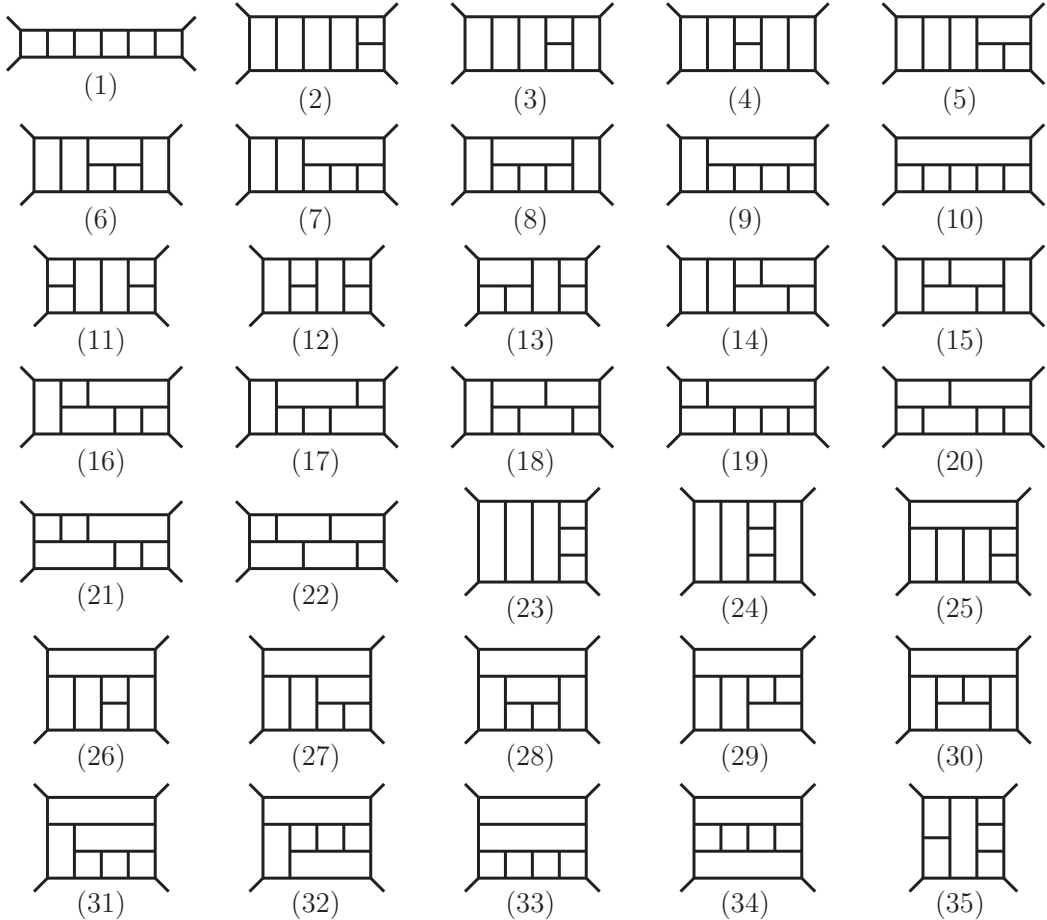}}
\caption[a]{\small Graphs 1 through 35 for the planar six-loop
four-point amplitude.}
\label{SixLoopPlanar1Figure}
\end{figure}

\begin{figure}[th]
\centerline{\epsfxsize 5.5 truein \epsfbox{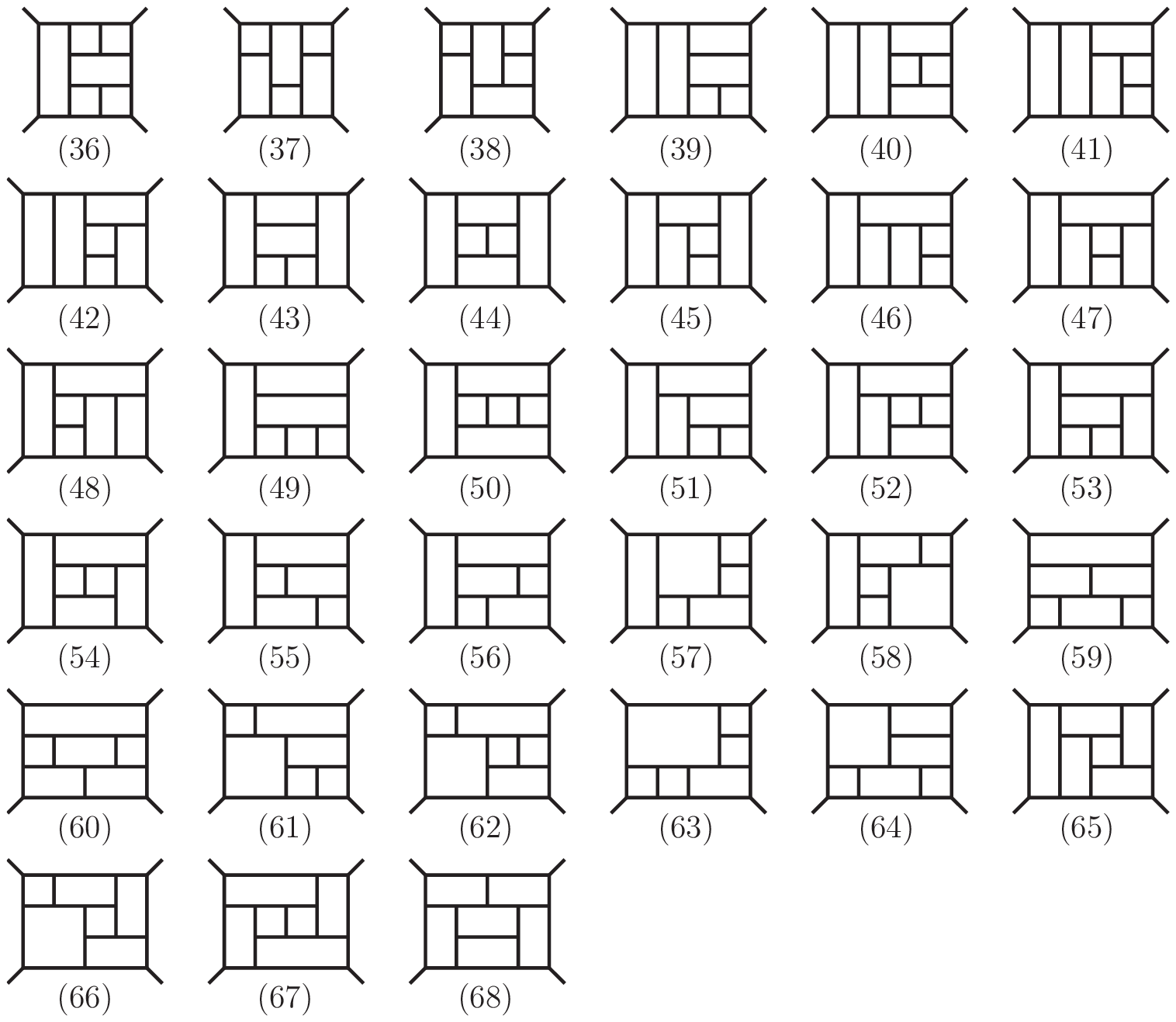}}
\caption[a]{\small Graphs 36 through 68 for the planar six-loop
four-point amplitude.}
\label{SixLoopPlanar2Figure}
\end{figure}

We write the Ansatz for the six-loop planar amplitude as
\begin{eqnarray}
A_4^{(6)}(1, 2, 3, 4) = i^6 s t A_4^\tree(1, 2, 3, 4)\, 
\int \prod_{l=5}^{10} {d^D p_l \over (2 \pi)^D} \, {\cal I} \,,
\label{SixLoopAmplitude}
\end{eqnarray}
where $A_4^\tree(1,2,3,4)$ is the color-ordered tree amplitude and the
Mandelstam invariants are $s = (k_1 +k_2)^2 $ and $t = (k_2 +k_3)^2$.
For bookkeeping purposes we organize the integrand in terms of graphs
with only cubic vertices.  We incorporate any contact (four-point)
interactions by including numerator terms that can potentially cancel
propagators.  Thus there is no loss of generality in using cubic graphs. 
We decompose the integrand ${\cal I}$ as 
\begin{equation}
{\cal I} = \sum_{D_4}\sum_{i=1}^{68} \frac{{\cal I}_i}{S_i} 
 = \sum_{D_4} \sum_{i=1}^{68} \frac{1}{S_i}
   \frac{N_i}{\prod_{\alpha_i=5}^{23} p_{\alpha_i}^2} \,.
\label{IntegralSum}
\end{equation}
The sum runs over a set of distinct planar cubic graphs, which
contribute in all eight possible arrangements generated
by the dihedral group $D_4$ (corresponding to symmetries of a square
with corners labeled by the four external momenta).
The dimension of the symmetry group leaving each diagram invariant is $S_i$.
At six loops, there are 68 non-vanishing topologically distinct graphs 
shown in \figs{SixLoopPlanar1Figure}{SixLoopPlanar2Figure}.
The product in the denominator of \eqn{IntegralSum} runs over the 19
internal Feynman propagators of each labeled graph.  The numerators,
$N_i$ of each integral are polynomials,
\begin{equation}
N_i = \sum_j a_{ij} M_{ij} \,,
\end{equation}
where the monomials $M_{ij}$ depend only on Lorentz invariants constructed
from the dual (loop) momenta for each diagram,
and the $a_{ij}$ are numerical coefficients (rational numbers)
to be determined from various constraints.  

As a first step, we require the monomials to have the proper weight
under dual conformal transformations.  To expose the dual conformal
properties we use the standard~\cite{MagicIdentities} dual variables
$x_i-x_j=x_{ij}$, with
\begin{equation}
 x_{41}=k_1\,, \hskip 0.5 cm x_{12}=k_2,\hskip 0.5 cm x_{23}=k_3,
\hskip 0.5 cm x_{34}=k_4\,, \
\end{equation}
where $k_i$ are the external momenta.
As discussed in detail in ref.~\cite{PlanarFive}, a practical way of
expressing the internal momenta of a diagram in terms of dual variables is to
use an $(L+1)$-particle cut, which divides the $L$-loop amplitude into two
tree amplitudes connected by $(L+1)$ cut legs.  At six loops, 
we consider a seven-particle cut in the $s = (k_1+k_2)^2$ channel.
The seven cut legs carry momenta $p_5, p_6, \ldots, p_{11}$.
The six dual loop momenta $x_5,x_6,\ldots,x_{10}$ are then defined by,
\begin{eqnarray}
x_{45}=p_5 \,,\hskip 0.5 cm  x_{56}=p_6 \,,\hskip 0.5 cm
x_{67}=p_7  \,,\hskip 0.5 cm x_{78}=p_8  \,,\hskip 0.5 cm
x_{89}=p_9  \,,\hskip 0.5 cm    x_{9, 10}=p_{10}\,.
\label{map}
\end{eqnarray}

The key dual conformal properties follow from the behavior of the integrand
under dual coordinate inversion, which maps
\begin{equation}
x^\mu_i \rightarrow 
\frac{x^\mu_i}{x^2_i}\,,~~~~~~x_{ij}^2 \rightarrow 
{x_{ij}^2 \over x_i^2 x_j^2}\,.
\end{equation}
In four dimensions, dual conformal invariance requires that each term
in the integrand scales as~\cite{MagicIdentities}
\begin{equation}
{\cal I}_{i}  \rightarrow \biggl( \prod_{j=1}^4 x_j^2 \biggr) \;
\biggl[ \prod_{l=5}^{10} (x_l^2)^4 \biggr] \ {\cal I}_i \,.
\label{IntegrandScaling}
\end{equation}
The integrands of planar MSYM in $D$ dimensions have been shown to
transform in exactly the same fashion to all loop orders, at least for
$D\le6$~\cite{SixDCheck,LoopConformal}.  This property is sufficient
for our purposes, since we are mainly interested in the integrand in $D=5$.

\begin{figure}[t]
\centerline{\epsfxsize 3truein \epsfbox{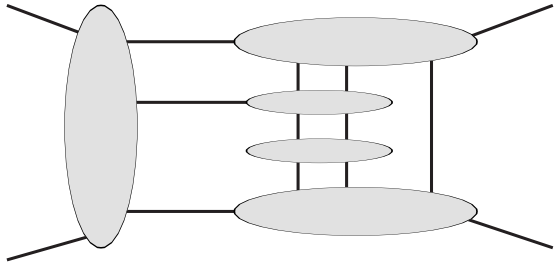}} 
\caption{The six-loop cut evaluated in $D=6$ in ref.~\cite{SixDCheck}.
}
 \label{SixLoopCutFigure}
\end{figure}

The $(L+1)$-particle cuts can also be used to generate the complete
list of graphs needed at six loops.  One considers all possible
sewings of two tree-level cubic graphs that appear in these
cuts~\cite{PlanarFive}.  (We modify the procedure slightly compared to
ref.~\cite{PlanarFive} by including only diagrams with cubic
vertices.) In principle there are dual conformal graphs with four- or
higher-point vertices that are not generated by the product of tree
graphs of the $(L+1)$-particle cuts; however, all such potential
contributions, including those not detectable in the $(L+1)$-particle
cuts, can be assigned to graphs with only cubic vertices by
multiplying and dividing by appropriate propagators.  The construction
of the potential numerators of each graph is then accomplished
conveniently using the dual-graph representation, which exposes the
dual conformal properties more simply.

Given a dual graph in the sewing, we construct the possible
monomials $M_{ij}$ as products of dual-momentum invariants $x_{ij}^2$.
We keep only those $M_{ij}$ terms with the dual conformal scaling dictated
by \eqn{IntegrandScaling}.  To determine the rational-number
coefficients $a_{ij}$ we use generalized unitarity.  
A large number of coefficients are easy to identify, essentially by
inspection, because the corresponding unitarity cuts are so simple.
In particular, contributions with either a two-particle
cut or a box subdiagram can be written down immediately, following the
discussion in refs.~\cite{BRY,CompleteFourLoopYM}.

In addition, many of the coefficients $a_{ij}$ vanish.
All but one of the vanishings
can be identified using the observation of ref.~\cite{KorchemskyZero}
that when the external momenta are taken off-shell the integrals must be
infrared-finite in four dimensions.  The sole integral with a vanishing
coefficient which cannot be identified in this way is the integral
displayed in Fig.~12 of ref.~\cite{PlanarFive}.  It consists of
two identical three-loop three-point integrals, containing only
box subdiagrams, and connected to each other by one common external leg.

The unitarity cuts include a sum over states in the supermultiplet
for each cut leg.  For generic cuts, the state sums are
straightforward to implement numerically in four or six
dimensions~\cite{StateSum,SixDCheck,SuperSum}.  We have evaluated all
four-dimensional cuts that decompose the amplitude into a sum of
products of three-, four- or five-point amplitudes, as well as a
variety of cuts involving six-point amplitudes. These cuts suffice to
uniquely determine all the $a_{ij}$.  The two-particle cut~\cite{BRY}
and ``box cut''~\cite{CompleteFourLoopYM} are valid in $D$ dimensions;
hence all contributions to the integrand that are visible in such cuts
are valid in any number of dimensions.  In addition, a rather
non-trivial cut of our expression, shown in \fig{SixLoopCutFigure},
was computed previously~\cite{SixDCheck} using the six-dimensional
spinor-helicity formalism of ref.~\cite{SixDHel} and superspace of
ref.~\cite{DHS}.  Thus this cut is valid for $D\le 6$. This provides a
highly nontrivial confirmation that the integrands are valid in $D=5$.

We have performed a variety of consistency checks on the integrand.
The unitarity cuts offer highly nontrivial self-consistency checks,
because the same monomial $M_{ij}$ can be visible in multiple cuts.
As already mentioned we compared our integrand result against that of
ref.~\cite{EdenRep}.  As a further non-trivial check we confirmed that
the unphysical singularities described in ref.~\cite{CachazoSkinner}
all cancel.

\begin{figure}[th]
\centerline{\epsfxsize 6.5 truein \epsfbox{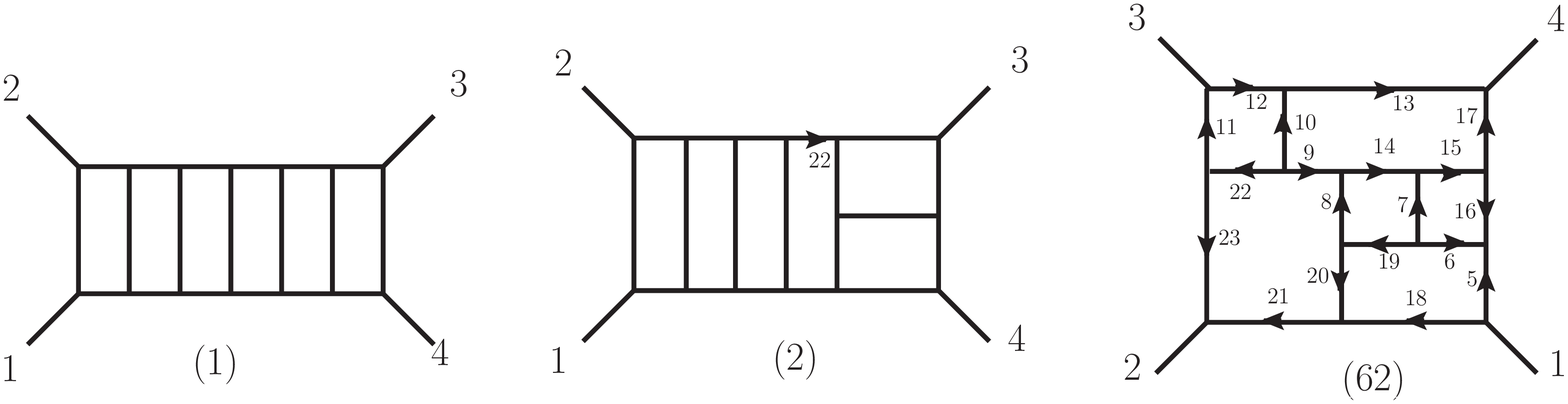}}
\caption[a]{\small A few sample graphs with labels corresponding 
to the labels in \eqns{SimpleSampleNumerator}{ComplexSampleNumerator} and 
in the ancillary file~\cite{AttachedFile}. The external momenta are outgoing.
}
\label{SampleGraphsFigure}
\end{figure}

Some of the numerators $N_i$ are quite simple.  For example, for
the graphs labeled (1) and (2) in \fig{SampleGraphsFigure}, they are
just
\begin{equation}
N_1 = s^5\,,  \hskip 3 cm 
N_2 = s^4 (p_{22} - k_3)^2\,.
\label{SimpleSampleNumerator}
\end{equation}
Other numerators are more complex. For example, the numerator of
graph (62) with the labels in \fig{SampleGraphsFigure} is
\begin{eqnarray}
N_{62}& =& \frac{1}{2} s t \, p_5^2 p_9^4 
      + t (k_1 + p_{21})^2 (k_2 - p_{18})^2 p_9^2 p_{17}^2 
      - t p_{17}^2 p_{20}^2 (k_1 + p_{21})^2 (k_3 + p_{13})^2 
     \label{ComplexSampleNumerator} \\
&&  \null 
      - s p_5^2 p_9^2 (k_1 + p_{17})^2 (k_3 + p_{23})^2
      - s p_{17}^2 p_{20}^2 (k_1 + p_{17})^2 (k_3 + p_{23})^2  \nn \\
&&\null
      - s t  p_9^2 (k_1 + p_{21})^2 (p_{20} -p_9)^2
      + t (k_1 + p_{21})^4 (k_3 + p_{13})^2 (p_{20} -p_9)^2  \nn \\
&&\null
      + s (k_1 + p_{17})^2 (k_1 + p_{21})^2 (k_3 + p_{23})^2 (p_{20} - p_9)^2
  \,.
\nn
\end{eqnarray}
In these expressions we have chosen labels that line up with the ones
in the ancillary text file~\cite{AttachedFile}.  The symmetry factors
of these graphs are $S_1=4$, $S_2=2$ and $S_{62}=1$.  The complete
sets of diagrams, numerators $N_i$, and symmetry factors $S_i$, are
included in the ancillary file~\cite{AttachedFile}.

\section{From the amplitude to vacuum diagrams}
\label{VacuumSection}

Because the UV divergences arise from integration regions in which the
loop momenta are parametrically much larger than the external momenta,
extracting the UV divergences is a much simpler task than integrating
the complete amplitude. We follow the same strategy as in our previous
papers~\cite{PreviousGravityUV,FourLoopGravity,CK4l}, based on Taylor
expanding the integrands in small external momenta and then integrating the
resulting vacuum integrals~\cite{Vladimirov, MarcusSagnotti}.  The
present case is relatively straightforward to analyze in the sense that the
six-loop amplitude contains no subdivergences in $D=5$, and because the
expected overall divergence is manifestly logarithmic.  However, the high
loop order makes the loop integration for the vacuum integrals highly
nontrivial.

\begin{figure}[th]
\begin{center}
\hskip -.5 cm {\epsfxsize 4.5 truein \epsfbox{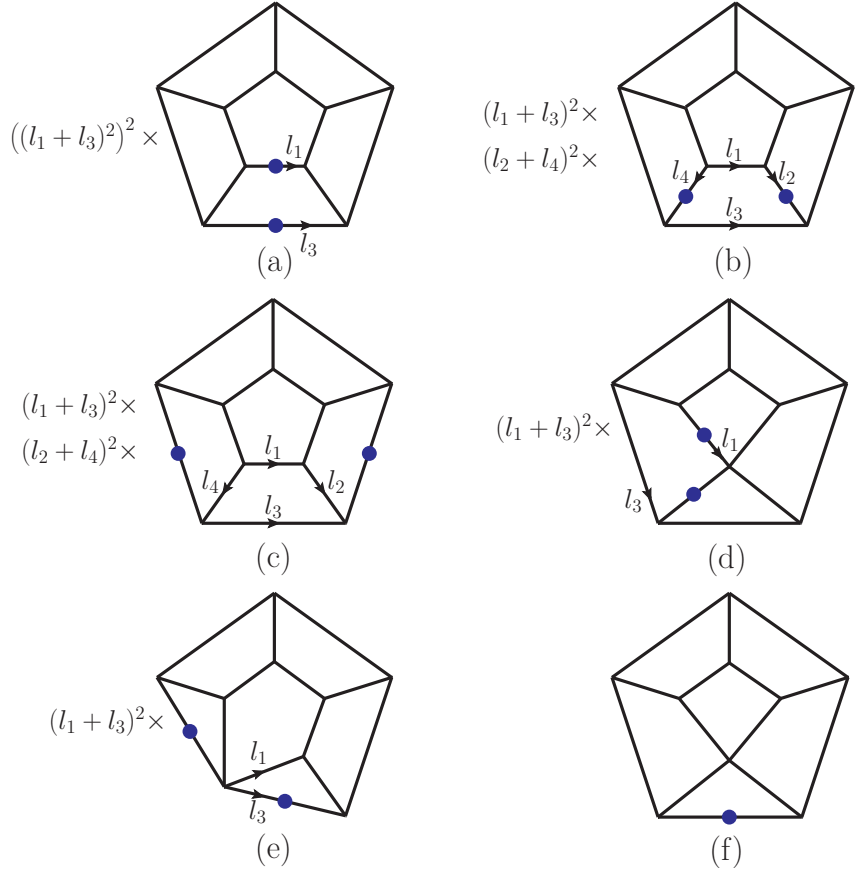}}
\caption[a]{\small The six distinct vacuum diagrams that appear in 
\eqns{UVdivergence}{eq:integrand}.  Each dot indicates that the 
corresponding propagator should be squared (doubled) in the integrand.
The five ``tensor'' integrals have numerator factors that are indicated
by the prefactors. The numerator factors are built from momentum
invariants involving a subset of the loop momenta, labeled by
$l_1,l_2,l_3,l_4$.}
\label{VacuumsFigure}
\end{center}
\end{figure}

\subsection{Obtaining the vacuum diagrams}

We consider the individual integrands ${\cal I}_i$ appearing in 
\eqn{IntegralSum} in the limit of small external momenta $k_j$, 
$j=1,2,3,4$.  We let $k_j\rightarrow \varepsilon k_j$,
and then expand in the small parameter $\varepsilon$, keeping only
the leading order.  This reduces each ${\cal I}_i$ to a sum over six
distinct vacuum integrands,
\begin{equation}
{\cal I}_i (\varepsilon k_j,p_l) \rightarrow \varepsilon^2  \hskip-4mm
 \sum_{x\in{\rm \{a,b,c,d,e,f\}}} \hskip-3mm 
 (s A_{i,x}+t B_{i,x}) {\cal V}^{(x)}(p_l) + {\cal O}(\varepsilon^3)\,,
\end{equation}
where $A_{i,x}$ and $B_{i,x}$ are rational numbers determined by the
expansion. (We will not list these coefficients separately for each diagram.)
After the above vacuum integrands ${\cal V}^{(x)}$ are integrated over
the six loop momenta $p_5,p_6,\ldots p_{10}$ in $D=5-2\epsilon$, with
the measure
\begin{equation}
\int \prod_{l=5}^{10} {d^{5-2\epsilon} p_l \over (2 \pi)^{5}} \,,
\label{FiveDmeasure}
\end{equation}
we obtain six vacuum integrals,
$V^{\rm (a)}, V^{\rm (b)}, \ldots, V^{\rm (f)}$, shown in \fig{VacuumsFigure}.
These integrals have numerator factors which are indicated to the left of
each graph, and either one or two doubled propagator, whose location
is indicated by a dot.
The integrals $V^{(x)}$ contain no subdivergences; each integral has a single
overall UV divergence in $D=5$ when all six loop momenta become large.
Hence the integrals have only simple poles in $\epsilon$.

Collecting the contributions from the 68 distinct integrals
in the six-loop amplitude \eqn{SixLoopAmplitude}, we obtain the following UV
divergence
\begin{equation}
A^{(6)}_4\Big|_\text{D=5,\, \rm div. }
 = 6 s t u A^{\tree}_4(1,2,3,4) ( V^{\rm (a)} + V^{\rm (b)} + 2 V^{\rm (c)}
 + 4 V^{\rm (d)} + 2 V^{\rm (e)}
 - 2 V^{\rm (f)} ) \,.
\label{UVdivergence}
\end{equation}
This simple expression for the divergence appears to be nonzero at
first glance. However, we should expect that these six integrals are
not independent so there may be cancellations among them.  While all
six integrals are positive definite after Wick rotation, $V^{\rm (f)}$
enters with a coefficient of the opposite sign from the others.  Thus
no conclusion can be reached as to whether this expression vanishes or
not without a careful analysis.

\begin{figure}[tb]
\centerline{\epsfxsize 1.6 truein \epsfbox{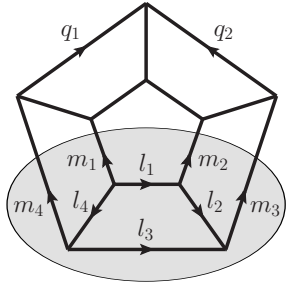}}
\caption[a]{\small The canonical vacuum integral.  All contributions
  can be expressed in terms of this diagram.  The differences between
  each contribution can be assigned to the one-loop subintegral
  indicated by the shaded region.}
\label{TargetVacuumFigure}
\end{figure}

We begin the analysis by collecting all integrals under one common
integration.  By multiplying and dividing by propagators, all contributions
can be brought to the form of the single vacuum graph displayed in
\fig{TargetVacuumFigure}.  Then by appropriately relabeling each term 
from the original $p_i$ to another set of loop momenta,
all contributions can be rearranged to differ in only a single
one-loop subintegral, indicated by the shaded region in
\fig{TargetVacuumFigure}.  The momenta $l_1, l_2,l_3, l_4$ are carried
by the indicated one-loop subintegral.  The momenta $m_1, m_2, m_3, m_4$
are the four momenta external to the one-loop box subintegral.  In
addition, we need two other independent loop momenta, which we take to
be $q_1$ and $q_2$, as indicated in \fig{TargetVacuumFigure}.  Of these
momenta, we take $l_1, m_2, m_3, m_4, q_1, q_2$ to be the six independent
ones.  In order to make the analysis easier, we may symmetrize the
contributions over the automorphisms of the vacuum integral.

After appropriate relabelings, we can express the sum over all vacuum
integrals as a single integral,
\begin{eqnarray} \label{eq:integrand}
V & \equiv & V^{\rm (a)} + V^{\rm (b)} + 2 V^{\rm (c)} + 4 V^{\rm (d)}
  + 2 V^{\rm (e)} - 2 V^{\rm (f)}
\nn \\ &=& 
\int \frac{d^{5 -2 \eps} l_1}{(2 \pi)^5} \, 
           \frac{d^{5 -2 \eps} m_1}{(2 \pi)^5} \,
           \frac{d^{5 -2 \eps} m_2}{(2 \pi)^5} \,
           \frac{d^{5 -2 \eps} m_3}{(2 \pi)^5} \,
           \frac{d^{5 -2 \eps} q_1}{(2 \pi)^5} \,
           \frac{d^{5 -2 \eps} q_2}{(2 \pi)^5} \;
 \frac{N_{\rm vac}} {l_1^2 l_2^2 l_3^2 l_4^2 m_1^2 m_2^2 m_3^2 m_4^2} \\
&& \hskip 1 cm 
\times \frac{1}{q_1^2 q_2^3 (q_1 + q_2)^2 (q_1 - m_4)^2  
  (q_1 - m_4 - m_1)^2 (q_2 - m_2 - m_3)^2 (q_2 - m_3)^2 }\,. \nonumber
\end{eqnarray}
where $\eps$ is the dimensional-regularization parameter and the
vacuum ``numerator'' is
\begin{eqnarray}
{N_{\rm vac}}&= &
 (l_1+l_3)^2 \biggl[ 
  \frac{(l_1+l_3)^2}{l_1^2 l_3^2}
 \ +\ \frac{(l_2+l_4)^2}{l_2^2 l_4^2}
\ +\ (l_2+l_4)^2\left( \frac{1}{m_1^2 m_2^2} + 
 \frac{1}{m_3^2 m_4^2}  \right)
\nonumber \\ && \hskip1.3cm\null
 + \left(
 \frac{ l_1^2 }{l_4^2 m_1^2} + \frac{ l_1^2 }{l_2^2 m_2^2}
 + \frac{ l_3^2 }{l_2^2 m_3^2} + \frac{ l_3^2 }{l_4^2 m_4^2} 
 \right) 
 + \frac{1}{2} \left(
   \frac{l_4^2}{l_1^2 m_1^2} + \frac{l_4^2}{l_3^2 m_4^2}
 + \frac{l_2^2}{l_1^2 m_2^2} + \frac{l_2^2}{l_3^2 m_3^2} \right)\biggr]
\nonumber \\ && \null
-\left(
\frac{l_1^2}{l_3^2} + \frac{l_3^2}{l_1^2}\right)
 \,,
\label{VacuumNumerator}
\end{eqnarray}
using the momentum labels in \fig{TargetVacuumFigure}; that is,
$l_2=l_1-m_2$, $l_3=m_2+m_3-l_1$, $l_4=-m_1-l_1$ and $m_4=-m_1-m_2-m_3$. 
By a slight abuse of convention, we call this a numerator even though it
is nonlocal.  The numerator depends only on the momenta internal and
external to the one-loop box subdiagram indicated by the shaded region
in \fig{TargetVacuumFigure}.

Because of the minus sign in the last term
--- corresponding to $-2V^{\rm (f)}$ in \eqn{UVdivergence} --- 
it is easy to see that this integrand is not positive definite, even after
Wick rotation.  For example, for $l_1=-l_3$ all terms but the last one
vanish, making the integrand negative; and in the region where $0<m_4^2$
is much smaller than all other momentum invariants, the terms with $1/m_4^2$
factors will dominate, making the integrand positive.  Thus, 
even after combining all contributions into a single integrand, there does
not appear to be a simple way to determine the positivity, or vanishing,
of the integral.

\subsection{Simplifying the vacuum integrals via consistency relations}

To simplify the expression further we need to identify
relations between the different vacuum integrals.
The integral identities
that we need are related to integration-by-parts identities
(IBP)~\cite{CTIBP}, although they are only valid for the
leading $1/\eps$ UV pole in the critical dimension, $D=5-2\eps$. These
{\it consistency relations}~\cite{CompleteFourLoopYM} are obtained by
demanding that different loop-momentum parameterizations of the
integrals lead to the same final results.  Using these relations we
can both simplify the UV divergence, and give important cross-checks
of the numerical evaluation.  The latter use is particularly
important, because the relations give independent estimates of the
numerical uncertainties. The integral consistency relations also 
offer a potential path to finding a positive-definite expression 
for the divergence.

We now sketch a derivation of the consistency
relations. For each of the 68 graphs describing the
amplitude, we have simple relations from the shift invariance
of the integrals,
\begin{equation} 
0= 
\frac{\partial}{\partial q_{m}^\mu}\int \prod_{l=1}^{6} \frac{d^D
  p_l}{(2\pi)^D}\, {\cal I}_i[\tilde N_i](k_j,p_l+q_l)\,,
\label{IBPdef}
\end{equation}
for each $q_m \in \{q_1,...,q_6\}$, and ${\cal I}_i$, for
$i=1,\ldots,68$, are the integrands of the distinct graphs. The
$\tilde N_i$ are general numerator polynomials in the momentum invariants
of the graphs.  These polynomials are chosen to generate useful
identities and are {\it not} the numerators $N_i$ of the amplitude.
A judicious choice of a large set of $\tilde N_i$ will lead to a large
set of linear relations for various vacuum integrals, which will include
(but will not be limited to) the desired integrals $V^{(x)}$.  Below we
describe such a judicious choice of numerators.

The identity~(\ref{IBPdef}) follows because the $q_l$ momentum
dependence of the integrands is completely removed by a change of
variables in the measure of \eqn{IBPdef}. In fact, \eqn{IBPdef} is
simply a statement that the integrals are reparametrization invariant
under constant shifts.

Next we expand the integrands in small external momenta,
$k_j\rightarrow \varepsilon k_j$, with $\varepsilon$ a small
parameter. In doing so, it is convenient to treat the integrands as
belonging to equivalence classes controlled by the reparametrization
freedom,
\begin{equation}
{\cal I}_i[\tilde N_i](\varepsilon k_j,p_l+q_l)\, 
\sim \,{\cal I}_i[\tilde N_i](\varepsilon k_j,p_l)\,.
\label{IntegrandEquiv}
\end{equation}
Expanding the two sides of \eqn{IntegrandEquiv} in $\varepsilon$ would
not yield any nontrivial equations, only trivial reparametrization
relations for vacuum integrals. However by combining the
reparametrization freedom and the small momentum expansion, that is,
by letting $q_l=\varepsilon \sum_j c_{lj} k_j$, we get nontrivial relations
between different vacuum integrals,
\begin{equation}
{\cal I}_i[\tilde  N_i]
\Bigl(\varepsilon k_j,p_l+\varepsilon \sum_j c_{lj} k_j \Bigr)\, \sim
\, {\cal I}_i[\tilde N_i](\varepsilon k_j,p_l)\,,
\label{IntegrandEquiv2}
\end{equation}
where $c_{lj}$ is an arbitrary integer-valued $6\times3$ matrix, where
different choices will generate different relations between vacuum
integrals.  The expansions in $\varepsilon$ of the two different
integrands will differ considerably. But by the reparametrization
freedom, the two sides must be equivalent as integrands, or equal
after integration.  Therefore the various vacuum integrals that arise
from integrating the coefficients of each element of $c_{lj}$ in
\eqn{IntegrandEquiv2} must satisfy nontrivial consistency relations.

It is important to make judicious choices for the integral numerators
$\tilde N_i$ used to generate useful consistency relations.  For example,
using the original $N_i$ appearing in the integrand of the amplitude
is not a good choice, because their divergences are manifestly
logarithmic in $D=5-2\eps$.  The UV divergence of a
logarithmically-divergent integral is given by the leading term in
$\varepsilon$.  This term is always insensitive to the shift in the
loop momenta $q_l$, and so the available consistency relations become
trivial.  The first nontrivial relations are obtained using numerators
$\tilde N_i$ containing one additional power of a loop momentum, which
give linearly-divergent integrals in $D=5-2\eps$. The relevant vacuum
integrals relations are obtained from the next-to-leading term in
$\varepsilon$, which will differ on both sides of \eqn{IntegrandEquiv}.
Numerators $\tilde N_i$ that give rise to quadratic divergences are
also useful for extracting integral relations. However, numerators
with an even higher degree are less helpful, since after taking
derivatives with respect to $\varepsilon$ they give rise to vacuum
diagrams with three or more doubled propagators.  These are outside
the class of integrals that we are interested in; the vacuum diagrams
in \fig{VacuumsFigure} have at most two doubled propagators.

Furthermore, one should not choose $\tilde N_i$ that give rise to
subdivergences in $D=5-2\eps$, because then the consistency relations
may be contaminated by relations that are only valid for overlapping
leading $1/\eps^n$ UV poles. Specifically, for integration in
$D=5-2\eps$ dimensions, UV subdivergences are possible in principle
for two- and four-loop subdiagrams. Any $\tilde N_i$ generating such a
subdivergence should be eliminated from the set of choices, because it
will not produce any useful identities.

\begin{figure}[tbh]
\centerline{\epsfxsize 4.5 truein \epsfbox{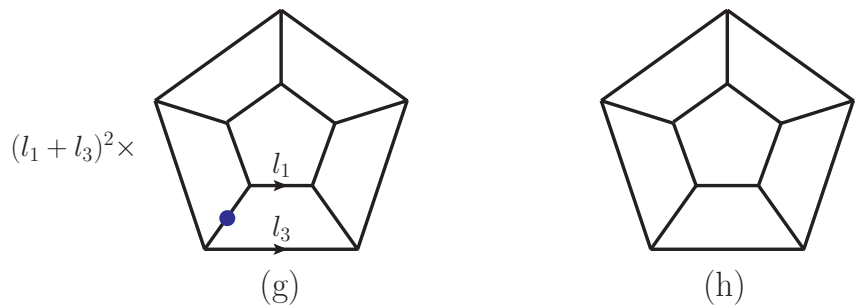}}
\caption[a]{\small Two simpler vacuum integrals that appear in the UV
  divergence after using integral identities. The third integral that
  enters the divergence is \fig{VacuumsFigure}(f).}
\label{SimplerVacuumsFigure}
\end{figure}

Generating a sufficient set of consistency relations then comes down
to varying the $\tilde N_i$ polynomials for an appropriately large
function space, without exceeding available computational resources.
This includes varying the matrix $c_{lj}$ that controls the
reparametrization of the integrand, and then, as explained, expanding the
integrals in small external momenta and demanding that the expansion is
consistent for different choices of $c_{lj}$.

After generating about 1000 independent consistency relations, we found
a much simpler three-term expression for the UV divergence of
the planar six-loop four-point amplitudes.  We then
generated an additional 7000 consistency relations, and no
further improvement was found.  Thus a search for identities beyond
the ones we found would probably be unfruitful, though we have not proven that
they do not exist.

\begin{figure}[tbh]
\centerline{\epsfxsize 4.5 truein \epsfbox{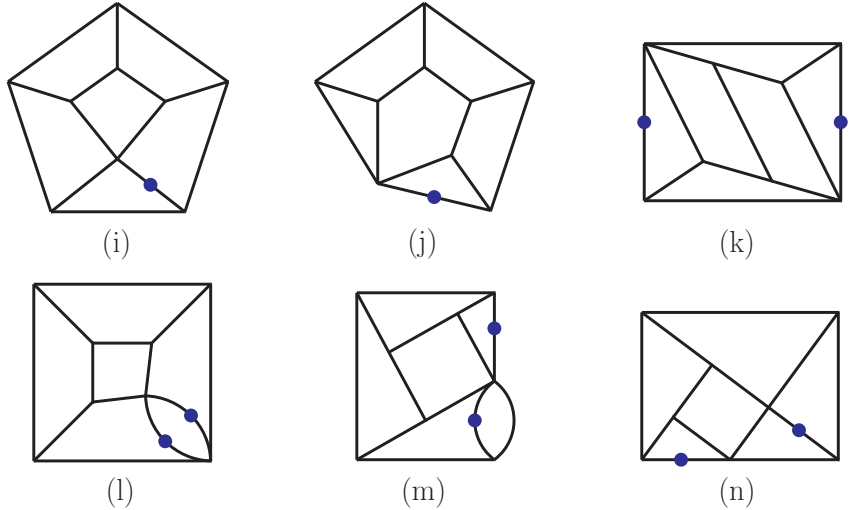}}
\caption[a]{\small Six scalar vacuum integrals that appear at
intermediate steps in integral identities that yield a simplified
UV divergence.}
\label{IntermediateVacuumsFigure}
\end{figure}

Using the derived consistency relations, the five tensor integrals 
$V^{\rm (a)},\ldots, V^{\rm (e)}$ can be reduced into eight scalar integrals
with no loop momentum in the numerator, plus one relatively simple integral
with only two powers of loop momentum in the numerator. 
One of the scalar integrals is $V^{\rm (f)}$ from \fig{VacuumsFigure}.
The eight new integrals
are shown in \fig{SimplerVacuumsFigure} and \fig{IntermediateVacuumsFigure}.
The most useful of the derived consistency relations are:
\bea
   V^{\rm (a)} &=& 2 V^{\rm (f)} + 2 V^{\rm (g)} - 4 V^{\rm (i)}
     + 2 V^{\rm (k)} - 2 V^{\rm (l)} \,, \nn \\
   V^{\rm (b)}&=& V^{\rm (f)} + 3 V^{\rm (g)} - 4 V^{\rm (i)}
     + 2 V^{\rm (k)} - 2 V^{\rm (l)} \,, \nn \\
   V^{\rm (c)} &=&  \frac{7}{2} V^{\rm (f)} - \frac{1}{2} V^{\rm (h)} 
     + V^{\rm (j)}  - 2 V^{\rm (m)}  + V^{\rm (n)}\,, \nn \\
   V^{\rm (d)} &=& \frac{1}{2} V^{\rm (f)} + 2 V^{\rm (i)}
     - V^{\rm (k)}  + V^{\rm (l)}\,, \nn \\
   V^{\rm (e)}&=& - V^{\rm (j)}  + 2 V^{\rm (m)}  - V^{\rm (n)}\,.
 \label{IBP}
\eea
%

\begin{figure}[tbh]
\centerline{\epsfxsize 1.2 truein \epsfbox{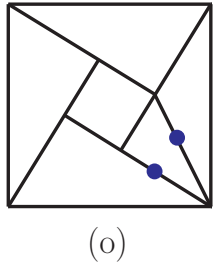}}
\caption[a]{\small  An extra scalar diagram used for checking integral
 identities.}
\label{ExtraDiagramFigure}
\end{figure}

After applying the integral identities~(\ref{IBP}) to
\eqn{UVdivergence}, we obtain the following simplified form for the
UV divergence,
\be
A^{(6)}_4\Big|_{D=5,\text{\, \rm div.}} =\ 6 s t u A^{\tree}_4(1,2,3,4)
       \bigl( 10 \, V^{\rm (f)} + 5 \, V^{\rm (g)} - V^{\rm (h)} \bigr) \,.
\label{SimpleUV}
\ee
Unfortunately, the coefficient of vacuum integral $V^{\rm (h)}$ has a
relative negative sign, so this simplified form is also not positive
definite. One may wonder if there exists a different choice of basis
of vacuum integrals that that allows the divergence to be expressed in
a positive-definite form. While this possibility cannot be excluded by
our analysis, we have not found such a representation, and it is quite
likely that \eqn{SimpleUV} is the simplest integral basis decomposition.

In addition to the needed relations, and as a cross check of the
numerical evaluation in \sect{ComputationalDetailsSection}, we offer
one very simple integral identity between three scalar integrals
\be
V^{\rm (o)} = \frac{1}{2} V^{\rm (f)} + V^{\rm (i)}\,,
\ee
where $V^{\rm (o)}$ is shown in \fig{ExtraDiagramFigure}.

\section{Review of sector decomposition}
\label{SectorDecompositionSection}

Given that even the simplified form~(\ref{SimpleUV}) still leaves open
the question of whether the amplitude diverges in $D=5$, and because
analytic techniques are not yet powerful enough to cope with generic
six-loop integrals, we have resorted to a numerical evaluation of the
relevant vacuum integral using the method of sector decomposition.
In this section we review this method, focusing on the salient
features needed in our calculation.  

If the amplitude under consideration is UV divergent, numerical
evaluation with even a modest (say 5\%) accuracy will suffice 
to prove it beyond any doubt.  Indeed, here we shall provide such a
numerical proof for MSYM in $D=5$, settling the question of the potential
finiteness of this theory.
Even had it turned out finite, numerical analysis can provide
important evidence in favor of this hypothesis.  We are
interested in the leading-logarithmic divergence of an integral
with no subdivergences.  After Feynman parametrization of the 15 propagators
and integrating over the loop momentum, the overall divergence appears
as a coefficient of a convergent $14$-dimensional parametric integral.
Eqs.~(\ref{UVdivergence})
or~(\ref{SimpleUV}) are fairly simple and it might seem to be an
easy job to estimate it by Monte Carlo methods.  If we estimate the
integral as an average over $N$ uniformly distributed samples, we
might imagine that the statistical uncertainty of such an estimate would
be $\sigma/\sqrt{N}$, where $\sigma$ is the standard deviation of the
integrand ($\sigma^2 \equiv \int d^Dx (f(x)-\bar{f})^2$ with $\bar{f}$ 
the average value).   An accuracy of $1\%$ would seem easily attainable.

However, the situation is more complex.  The problem is that
Feynman integrals in general, including \eqns{eq:integrand}{SimpleUV},
are not sufficiently convergent because of
endpoint singularities.  UV divergences in themselves are not a major
problem, and are usually dealt with by subtracting a simpler integrand
with the same divergent behavior.  In the case at hand, even this is
not necessary because the coefficient of the divergence is given by an 
(absolutely) convergent integral.  However, the
square of this integrand is not integrable.  Because of this, the
variance $\sigma^2$ will diverge, and Monte Carlo estimation with the
uniform measure has uncontrolled errors.  In practice, it does not
work.

\subsection{Sector decomposition overview}
A straightforward way to deal with this problem is to
carry out Monte Carlo integration with a sampling measure designed to 
overcome this problem.  Let $d\mu(x)$ be the sampling measure.
Then we rewrite
\bea
I &=& \int d^Dx\ f(x)\nn \\
 &=& \int d\mu(x) \left|\det\frac{d^D\mu(x)}{d^Dx}\right|^{-1}\ f(x)\nn \\
 &\equiv & \int d\mu(x)\ F(x)\,,
\eea
where $F(x)$ combines the original function $f(x)$ with a change of
measure factor.  A particular case would be to take a new set of
variables $y$ functionally related to $x$, and $d\mu$ to be uniform
measure $d^Dy$, in which case this factor would be the Jacobian for
the change of variables.  The error estimate for a Monte Carlo integral 
with this sampling measure is $\sigma_F/\sqrt{N}$ where $\sigma_F$ is the
new standard deviation,
\be
\sigma^2_F = \int d\mu(x)\ \left( F(x) - \bar F \right)^2 \,,
\ee
where $\bar F$ is the average value of $F$.  By reducing the variance,
one can improve the accuracy of the Monte Carlo.  In principle one
could change variables to absorb all of the variation of $f(x)$ into
the measure to make $F(x)$ constant, in which case the exact result
would be obtained from a single sample.  Of course, in practice such a
change of variables would be prohibitively difficult.

For an integrand with power-like singularities, such as
\be \label{eq:changevar-power}
\int dx\, x^{\alpha-1} (1 + f(x)) \,,
\ee
we could use the change of variables $y = x^{\alpha}/\alpha$, which
absorbs the singularity into the measure and does not much complicate
the integrand.  The problem with applying this to a multi-dimensional
Feynman integrand is that it has many different power-like singularities,
arising from the many orders in which one can take the different parameters
$x_i$ to zero.

The solution is to decompose the integration region into subregions or
``sectors,'' each of which has at most one singular behavior of this
type.  If we can do this, we can apply the change of variable
\eq{changevar-power}, or its multivariate generalization, to regularize
the integral in each sector.  While this approach was long used in
formal proofs of perturbative renormalizability \cite{Hepp, Speer,
  Breitenlohner}, it seems to have been first used in numerical
integration by Binoth and Heinrich~\cite{BinothHeinrichSector}.  Since then
has been implemented in several computer packages for numerically
evaluating Feynman diagrams, starting with 
Bogner and Weinzierl~\cite{BognerWeinzierl} and including
refs.~\cite{CarterSecDec,Fiesta,Fiesta2}.  Let us explain the basic ideas,
leaving the details specific to our computation to
\sect{ComputationalDetailsSection}.

We begin with an elementary example (from section 2 of
ref.~\cite{HeinrichReview}), the two-dimensional integral
\be I = \int_0^1 dx \int_0^1 dy \frac{x^\alpha y^\beta}{x + (1-x) y} \, .
\ee 
The form of the denominator makes the limit $x,y\rightarrow 0$ hard to
control.  Although in this simple example we could change variables to
(say) $x$ and $w=x + (1-x) y$, this option will not be available for
more complicated integrands.

Rather, we split the integration region into two parts, region 1 with
$x\ge y$ and region 2 with $y\ge x$.
In region 1, we can make the change of variables
\be
x = x'\,, \qquad y = x' t' \,,
\ee
turning the integration region into $0\le x',t'\le 1$.
Similarly, in region 2 we take 
\be
x=y' t'\,, \qquad  y=y' \,,
\ee
again turning the integration region into a square.
The integral becomes (suppressing the primes)
\be
\label{eq:example-int}
I = \int_0^1 dx \int_0^1 dt \, \frac{x^{\alpha+\beta} t^\beta}{1 + (1-x) t}\,
+\, \int_0^1 dy \int_0^1 dt \, \frac{t^\alpha y^{\alpha+\beta}}{1+(1-y)t} \, .
\ee
Now the nature of the singularity is manifest in the leading monomial
terms, because the complicated denominator goes to $1$ in the singular
region.

The same idea can be applied to a function of $N$ variables, call
these $x_i$ with $i\in [1,N]$.  The integration region is decomposed
into $N$ subregions labeled by $a\in [1,N]$ and defined by the
inequality 
\be
 x_a \ge x_i\,, \qquad i\ne a \,.  
\ee
 In the $a$'th sector
we redefine 
\be \label{eq:mul-blowup} 
x_i = x_a x'_i\,, \qquad a\ne i \,,
\ee 
to turn the subregion back into a unit cube.  This will allow
pulling out an overall singular behavior controlled by $x_a$.  Of
course, the resulting integrand might still have a complicated
singularity in the other variables.  This must be dealt with by
iterating the procedure and dividing the subregion into
further subregions. Mathematically, this operation is called
``blowing up'' the singularity. 

In these subsequent subdivisions, one need not include all $N$
variables in the blowup; one can instead take a subset of the
variables and apply the same procedure.  These choices might be used
to simplify the result, or might even be needed in order for the
procedure to terminate.  The hope is that by choosing an appropriate
sequence of these operations, one can find a finite set of subregions
in each of which the integrand takes a simple form, as in
\eq{example-int}.

This procedure may be familiar to some readers from its use in
algebraic geometry and string compactification, and the following
remarks are addressed to them.  In complex algebraic geometry, one can
blow up an arbitrary point $p$ in an $N$-dimensional space, replacing
it with a $\IC\IP^{N-1}$.  This is done by taking coordinates in which
$x_i(p)=0$ and applying the same changes of variables; the $a$'th
subregion corresponds to the coordinate patch on $\IC\IP^{N-1}$ in
which we can take $x_a=0$.

Suppose that the integrand is a rational function with denominator
$D(x)$.  The singularity is then the set of all points satisfying
$D(x)=0$.  Let us denote this singular set as $\Delta$.  Since for a
Feynman integrand the function $D(x)$ is a polynomial, the set
$\Delta$ is by definition an algebraic variety, meaning the set of
solutions of a system of polynomial equations.  In fact it is a
hypersurface, since we are setting a single polynomial to zero.  In
this context, a natural thing to try is to blow up the space $\IC^N$
containing $\Delta$, to a variety $\pi:X\rightarrow \IC^N$ so that
$\pi^{-1}(\Delta)$ is nonsingular, meaning that it is defined locally
by a single equation $f=0$ with $\partial f\ne 0$.  If we can do this,
then the singular behavior of the integrand will simply be $1/f$ (or
perhaps some power of this).  By taking $f$ to be a local coordinate,
we would accomplish our goal of realizing the singular behavior in a
particularly simple form.

For present purposes, the main result of this mathematics is 
Hironaka's theorem on resolution of singularities, which states that
any singularity of an algebraic variety can be resolved (made nonsingular)
by a succession of blowups.  
Furthermore, there are algorithms for concretely
finding the resolution.  Thus, we can use a blowup algorithm
to resolve the singular locus $\Delta$, providing the multiparameter
generalization of \eq{example-int}.

While in ref.~\cite{BognerWeinzierl} this idea was used to give
blowup algorithms which are guaranteed to terminate, these algorithms
tend to produce a number of subregions which are exponential in the
number of variables --- this is perhaps natural as the number of
orderings of $N$ variables is $N!$.  This number of subregions would
be computationally infeasible for $N=15$.  Bogner and Weinzierl
\cite{BognerWeinzierl} also proposed a simpler heuristic algorithm
which, while not guaranteed to terminate, produces a simpler solution
when it does.  Another heuristic algorithm was proposed by Smirnov and
Tentyukov \cite{Fiesta}, which we now describe.

\subsection{Heuristic sector decomposition}

We recall from textbooks~(e.g.~ref.~\cite{IZ})
that the denominator of the Feynman integrand for a vacuum integral
is the $D/2$ power of the  Kirchoff polynomial of the graph,
\be \label{eq:kirchoff}
U_{\Gamma}(x_i) = \sum_T \prod_{i\notin T} x_i \,,
\ee
where the sum is taken over the spanning trees $T$ of the graph $\Gamma$.
We are interested in the limiting behavior of $U$ as
combinations of the variables go to zero.  This behavior is encoded in
its Newton polytope.  Let $\deg$ be the degree of a monomial,
considered as a vector in $\IZ^N$, so that
\be
\deg x_1^{n_1} x_2^{n_2} \ldots x_N^{n_N} \equiv (n_1,n_2,\ldots,n_N) \,.
\ee
The Newton polytope of $U$ is the convex hull of the degrees of each
of its terms; in other words, it is the set of all points in $\IR^N$ which
can be obtained as linear combinations of these degrees with
non-negative coefficients.

We need to desingularize each limit which takes a subset of the
variables to zero.  We now assume that $U$ is a polynomial with no
constant term, so that every monomial in $U$ will go to zero for some
such limits.  However, many of the monomials are subleading and do not
control any limit: if a monomial $M_1$ is the product of another
monomial $M_2$ with a monomial of non-negative degree, it is
subleading.  In terms of the degrees, this requires
\be \label{eq:degree-ineq}
\deg M_1 - \deg M_2 \ge 0\,,
\ee
for every component.

We refer to the points $\deg M_1$ which do {\it not} satisfy \eq{degree-ineq} 
for any $\deg M_2 \ne \deg M_1$ as the ``low points'' of the polytope.
If the polytope has a single low point, then by factoring out the 
corresponding monomial, one obtains a polynomial with a nonzero constant term.

If there are multiple low points, we need to do a blowup.  A blowup on
a subset $S$ of the variables subdivides the current sector into $|S|$
sectors.  In the $a$'th sector one applies the change of
variables~(\ref{eq:mul-blowup}).  This change of variables
operates on the Newton polytope as
\be
v \rightarrow v + e_a (\chi_S-e_a,v) \,,
\ee
where $\chi_S$ is the vector whose components are $1$ for $i\in S$ and
$0$ for $i\notin S$, and $(v,w)=\sum_i v_i w_i$.  We will then be able
to factor out a common monomial; in other words shift the entire
polytope in a way that keeps it in the upper quadrant.  If the result
includes the origin, we are done with this sector; otherwise we apply
the same procedure recursively.

The next problem is to decide which subset of variables to involve in
the blowup.  The goal is to eliminate as many low points as possible;
however it is better to leave the variables which do not contribute to
this goal out of the blowup, in order to maximize the degree of the
monomial that can be factored out.  A condition on the subset $S$
which favors this is to project the polytope onto the subspace $V_S$
and require that the low points of the projected polytope linearly
span this subspace.  The heuristic algorithm is simply to choose,
from the subsets satisfying this condition, the subset $S$ that has
the largest value of $\sum_{i\in S} i$.

This heuristic algorithm
is too simple to desingularize general polynomials,
including examples that are given in ref.~\cite{BognerWeinzierl}.
If applied to these examples, it will go into an infinite 
loop.\footnote{The ``Strategy X'' of ref.~\cite{BognerWeinzierl} and 
``Strategy S'' of ref.~\cite{Fiesta2} are somewhat more sophisticated and
can handle these cases.}  However, the heuristic algorithm
works for the class of polynomials in \eq{kirchoff}; that is,
scalar vacuum integrals with no doubled propagators.  It also turns out
to work for vacuum diagrams with no IR divergences and simple
numerator factors, including the integrand~(\ref{eq:integrand}).
As observed in ref.~\cite{SmirnovSectorDecomp}, it works because it
reproduces the results of a canonical sector decomposition procedure,
which associates sectors with maximal
forests~\cite{Speer,Breitenlohner,SmirnovBook}.  While we leave the
details for the references, a maximal forest is a hierarchical
decomposition of the graph into a set of subgraphs satisfying certain
conditions (each pair of subgraphs $(\gamma,\gamma')$ must obey one of
the relations $\gamma\subset\gamma'$, $\gamma'\subset\gamma$ or
$\gamma\cap\gamma'=\emptyset$ and $\gamma\cup\gamma'$ can be
disconnected by removing a single vertex).  It can be shown that a
maximal forest for a diagram with $L$ loops and $E$ edges contains $E$
trivial subgraphs (single lines) and $L$ nontrivial subgraphs, and the
associated sector involves $L$ blowups each on distinct variables.

If the heuristic algorithm is reproducing this decomposition, then 
since every sector involves a succession of $L$ blowups on distinct
variables, the algorithm is guaranteed to terminate with at most
$N!/(N-L)!$ sectors.  For $N=15$ and $L=6$ this is $3603600$ which is
not much larger than the actual numbers we obtained.  For $N=18$ and
$L=7$ it is about $1.6\times 10^8$.

\section{Numerical results and computational details}
\label{ComputationalDetailsSection}

\subsection{Numerical integration}

There are several software packages for carrying out sector
decomposition and subtraction of divergences, and for integrating the
resulting expressions numerically \cite{CarterSecDec,Fiesta,Fiesta2}.
We used the package {\sc FIESTA 2}~\cite{Fiesta2}, written in a combination
of Mathematica~\cite{Mathematica} and C++~\cite{Cpp}, and which can
take advantage of multiple processors.  However, a six-loop diagram is
too complicated to evaluate directly.  For example, the sector
decomposition (which is not very parallelized) takes about a day per
primary sector to run in Mathematica on a modern computer, and
produces a total code for the integrands which takes up hundreds of
gigabytes.  Thus the computation must be split into smaller parts to
make it feasible.

To do so, we only need a small portion of the {\sc FIESTA 2} software, and
we have extracted this portion and adapted it to our purpose by hand.
We did the sector decomposition on a small (10 node) cluster, and then
performed the numerical integrations on a large (1000 node) cluster, using
the adaptive quasi-Monte Carlo integrator \Vegas\ \cite{HahnCuba}.  An
important element in {\sc FIESTA 2} is the {\sc CIntegrate}
package (see ref.~\cite{Fiesta2}, appendix F), which accepts a symbolic
algebraic expression of the sort that can be produced easily by
Mathematica, and compiles it into pseudocode that can be executed
efficiently in C.  Using this package, we were able to break down the
computation by having Mathematica produce a file for each sector containing
its integrand, which could be passed to a \Vegas\ integration program.
The integrals are of course completely independent, so this step is easy to
parallelize.  This approach also
allowed us to keep the partial results for every sector, which helped
in debugging and uncertainty analysis.  The main cost was the need for
1--2 TB of disk storage, which is not large these days.

A rough estimate of the total running time can be obtained by
multiplying the number of sectors (about $10^6$ here) by the time to
integrate a sector, divided by the number of nodes.  With a regular
(low variance) integrand, the uncertainty as a function of the number
$N$ of samples goes somewhere between $N^{-1/2}$ for Monte Carlo and
$N^{-1}$ for quasi-Monte Carlo in low dimensions.  \Vegas\ did not
need more than $50,000$ samples to achieve our requested relative
precision of $10^{-3}$ in any sector, and this took 2-3 seconds to do.
Thus a million sector integration took less than an hour on the
cluster.

There were a number of reasons that this success was not guaranteed
from the start.  Even once we knew that we had of order $10^6$
sectors, the next possible pitfall was that the integral might be
small due to cancellations between larger results in individual
sectors.  The relative uncertainty of course depends strongly on the
relative signs of the intermediate results; in the best case (a single
sign) we might hope to gain a further statistical
$1/\sqrt{N_{\rm sectors}}$, while in the worst case the result might be
comparable to the largest statistical uncertainty in a single sector
(which could of course be improved by taking more samples) or even an
uncontrolled systematic uncertainty.  In fact, it appears that, at
least in our computations, sector decomposition does lead to
significant systematic uncertainties, as we will see below.

\begin{figure}[tb]
\centerline{\epsfxsize 5.0 truein \epsfbox{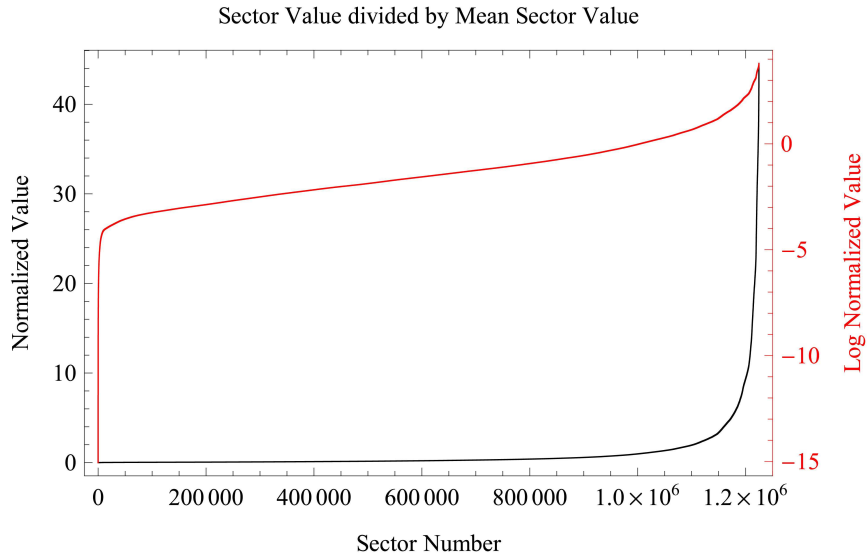}}
\caption{The sector integrals for diagram $V^{\rm (g)}$, in order of
increasing numerical value.  The dark bottom 
(black) line gives the sector value,
normalized by dividing by the average sector value. The  top light (red) line
gives the natural logarithm of the normalized sector value.}
\label{datalogplotFigure}
\end{figure}

\begin{figure}[tb]
\centerline{\epsfxsize 5.0 truein \epsfbox{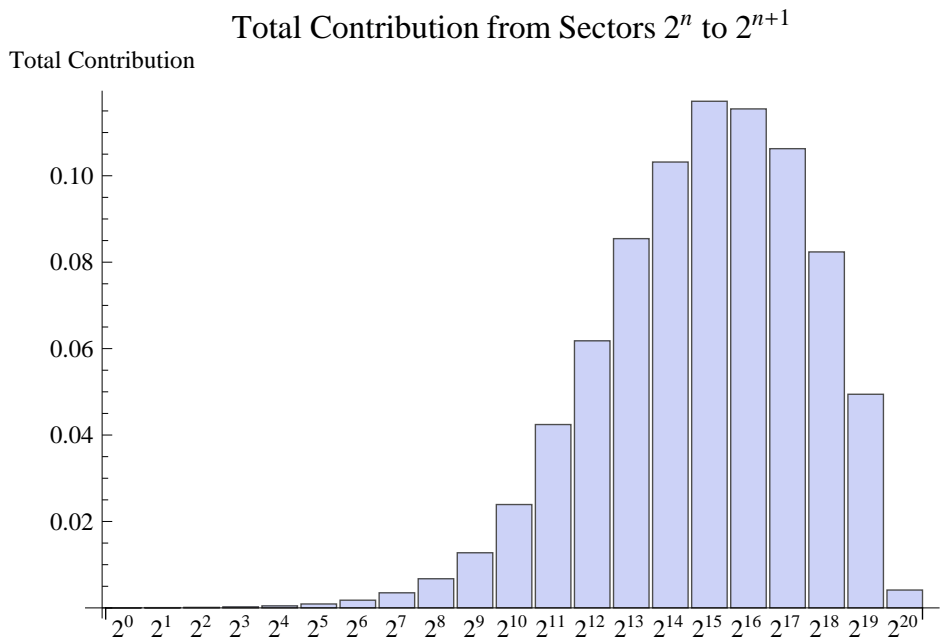}}
\caption{The total contribution of sectors $2^n$ through $2^{n+1}$,
ranked in order of decreasing numerical value. 
 }
\label{barplotFigure}
\end{figure}

The actual situation is best judged by looking at the partial results,
which we graph for diagram $V^{\rm(g)}$ in \fig{datalogplotFigure}.
There are $1224600$ sectors, all of which gave positive contributions.
The same is manifestly true for $V^{\rm (f)}$ and $V^{\rm (h)}$ (since
these integrands were positive), so there is no problem with
cancellations.

Although a few sectors give anomalously large values, the largest
sector value is only about $40$ times the average sector value, while
the largest uncertainty is about $180$ times the overall standard
deviation.  One also sees that the distribution of sector integrals is
roughly exponential except at the ends.  This point is made in a
different way in \fig{barplotFigure}. In this plot the sectors are
listed in decreasing numerical order, with the graph showing the total
contribution of each subset which has an exponentially growing number
of sectors.  Because the totals in each subset are roughly constant, this shows
that the sector contributions decrease exponentially.
Thus, while one cannot
reduce the calculation to evaluating a smaller number of dominant
sectors, there is structure which might be exploited to speed up the
computation.  For example, if one simply fits the distribution to an
exponential and integrates that, one gets within $20\%$ of the actual
total.

All this seems (at least naively) consistent with the uncertainties being
entirely statistical, in which case it would be appropriate to add
them in quadrature to get the statistical uncertainty estimate for
integral $V^{(g)}$ of $6 \times 10^{-4}$\%.
However, in previous use of sector decomposition, it has
been observed that a purely statistical combination underestimates the
true uncertainty.  A more conservative estimate would be to treat the
uncertainties as 100\% correlated --- what we will refer to as
a ``systematic'' uncertainty estimate.  In this case, we should
add the uncertainties in every sector.  For integral $V^{(g)}$,
this error is $0.29$\%.

In Table~\ref{NumericalResultsTable}, we collect our results for the
various vacuum integrals defined in \sect{VacuumSection}, along with
uncertainties that have been estimated by assuming 100\% correlation
among sectors.  These results can be used to confirm the consistency
relations~(\ref{IBP}); the degree of agreement is in turn a cross check
of the integration uncertainties.

For example, consider the first equation in \eqn{IBP}.  This relation
evaluates to the numerical values
\be
1.3958 \simeq 2 \times 0.7631 + 2 \times 0.8183 - 4 \times 0.5967
+ 2 \times 1.1493 - 2 \times 0.8391  = 1.3964,
\ee
which is certainly acceptable and consistent with the systematic uncertainty
hypothesis.  The other consistency relations in \eqn{IBP} work to a
similar accuracy.  Thus the combination of sector decomposition with
the \Vegas\ adaptive numerical integration appears to introduce
systematic uncertainty, in the sense that the error is correlated between
different sectors.  Presumably this has to do with the modeling of the
integrand at the sector boundaries.

Taking the uncertainties as 100\% correlated and conservatively
adding them directly (instead of in quadrature), the final result for
the UV divergence is
\bea
A^{(6)}_4\Big|_{D=5,\text{\, \rm div.}} &=& 
- {\cal X}\, 6 \, \Bigl( 10 \, V^{\rm (f)} + 5 \, V^{\rm (g)} - V^{\rm (h)} \Bigr)
         \nonumber \\
&=& - \frac{1}{\eps}  \frac{\cal X}{(4 \pi)^{15}}
  \, 6 \, \Bigl[ 10 \times (0.7631 \pm 0.0015)
                + 5 \times (0.8183 \pm 0.0024)
     \nn \\
 &&  \null \hskip 1.8 cm 
  -(0.2762 \pm 0.0008) \Bigr] \nn \\
&=& - \frac{1}{\eps}  \frac{\cal X}{(4 \pi)^{15}} (68.68 \pm 0.17) \,,
\label{SixLoopValue}
\eea
where 
\begin{equation}
{\cal X} = -s t u A_4^{\rm tree}(1,2,3,4) \,.
\label{CalKDef}
\end{equation}
It is clear from this result that the coefficient of the UV divergence 
is nonzero, well within the integration uncertainty.   This proves
that MSYM is perturbatively ultraviolet divergent in $D=5$.

\begin{table}[tb]
\def\hs{\hskip .7 cm $\null$}
\begin{center}
\begin{tabular}{cccc}
 \hs Integral\hs & \hs Value \hs &\hs Uncertainty
 \hs &  (Value) / $V^{\rm (h)}$\\
\hline
$V^{\rm (a)}$ & 1.3958 & 0.0043  & 5.05\\
$V^{\rm (b)}$ & 1.4522 & 0.0079  & 5.26\\
$V^{\rm (c)}$ & 1.3346 & 0.0069  & 4.83\\
$V^{\rm (d)}$ & 1.2643 & 0.0029  & 4.58\\
$V^{\rm (e)}$ & 1.1935 & 0.0026  & 4.32\\
$V^{\rm (f)}$ & 0.7631 & 0.0015  & 2.76\\
$V^{\rm (g)}$ & 0.8183 & 0.0024  & 2.96\\
$V^{\rm (h)}$ & 0.2762 & 0.0008  & 1\\
$V^{\rm (i)}$ & 0.5967 & 0.0012  & 2.16\\
$V^{\rm (j)}$ & 1.1490 & 0.0020  & 4.16\\
$V^{\rm (k)}$ & 1.1493 & 0.0019  & 4.16\\
$V^{\rm (l)}$ & 0.8391 & 0.0015  & 3.04\\
$V^{\rm (m)}$ & 1.8600 & 0.0028  & 6.74\\
$V^{\rm (n)}$ & 1.3755 & 0.0022  & 4.98\\
$V^{\rm (o)}$ & 0.9790 & 0.0017  & 3.55\\
\end{tabular}
\caption{The data used to numerically verify the integral consistency
  relations, as a means of assessing uncertainties in the
  numerical integration.  The columns labeled by ``Value'' and 
``Uncertainty'' are multiplied by $\eps (4\pi)^{15}$.
 \label{NumericalResultsTable} }
\end{center}
\end{table}

\begin{figure}
\centerline{\epsfxsize 4.2 truein \epsfbox{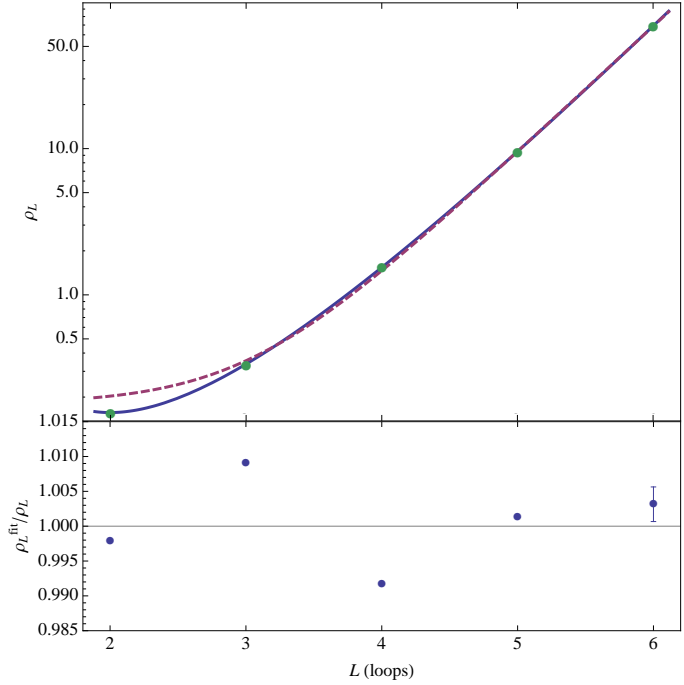}}
\vskip -1 cm 
\caption{The dots indicated the numerical coefficients in
  \eqn{LowerLoopValues} of the UV divergences in the critical
  dimensions.  The solid (blue) line is the result of fitting the
  parametric form in \eqn{SimplePrediction2} to the displayed results
  for $L=2,3,4,5,6$.  The dashed (purple) line is a fit to the
  parametric form in \eqn{SimplePrediction}. The lower panel shows
  that the relative error between the points and the fit in
  \eqn{SimplePrediction2} is within 1\%.}
\label{SixLoopFitFigure}
\end{figure}

\subsection{Extrapolating in loop order} 

Maximally supersymmetric Yang-Mills four-point amplitudes have a
smooth analytic behavior as a function of dimension, through at least
six loops.  As seen from their explicit forms the only dependence on
the space-time dimension is in the loop momentum integration measure.
Might this property lead somehow to a simple functional form
describing the numerical values of the divergences in the critical
dimensions, as a function of the number of loops $L$?  To check this
hypothesis, we plot the known values of the divergences in
\fig{SixLoopFitFigure} represented by the dots.  In doing so we extract some
simple overall factors, defining the numerical constant $\rho_L$ by
\begin{equation}
A^{(L)}_4 \bigr|_{D = 4+6/L,\, \rm div.}
 \ =\ {1\over\eps} \frac{(-1)^{L-1} \, {\cal X}}{(4 \pi)^{2 L +3}}
  \, \rho_L \,, 
\label{DefineRho}
\end{equation}
where ${\cal X}$ is defined in \eqn{CalKDef}.

The $L=1$ value is not plotted because it does not obey the critical
dimension bound given in \eqn{eq:FinitenessBound}, and
its divergence in $D=8$ differs kinematically from \eqn{DefineRho}
by a factor of $1/u$:
\begin{equation}
A^{(1)}_4 \bigr|_{D=8,\, \rm div.} = 
     \frac{1}{\eps}\frac{{\cal X}}{(4 \pi)^4}
     \, \frac{1}{6} \, \frac{1}{u}  \,.
\label{OneLoopValue}
\end{equation}
The values of the divergences in $D=4+6/L$ from two through five loops
have been given previously~\cite{BDDPR,CompactThree,CK4l,FiveLoopNew}.
Correcting a couple of overall signs, they are
\begin{eqnarray}
\rho_2 &=& \frac{\pi}{20} \,,   \nn \\
\rho_3 &=& \frac{1}{3} \,, \nn \\
\rho_4 &=& 6 \, \biggl[ \frac{512}{5} \Gamma({\textstyle \frac{3}{4}})^4 -
    \frac{2048}{105} \Gamma({\textstyle \frac{3}{4}})^3
    \Gamma({\textstyle\frac{1}{2}}) \Gamma({\textstyle\frac{1}{4}}) \biggr]
  \simeq 1.553 \,,  \hskip 1 cm \nn \\
\rho_5 &\simeq& 9.537 \,,
\label{LowerLoopValues}
\end{eqnarray}
where the expressions through $L=4$ are exact. The $L=5$ expression
is approximate, but it is accurate to the digits given.  

The linear behavior beyond $L=2$ in the upper panel of
\fig{SixLoopFitFigure} make clear that the coefficients of the
divergences have an approximately exponential behavior.  This
observation motivates a simple Ansatz for the approximate form of
the divergences at any loop order $L\ge 2$,
\begin{equation}
\rho_L \simeq b_1 c_1^{L+a_1/L} \,. 
\label{SimplePrediction2}
\end{equation}
The solid curve in
the upper panel of 
\fig{SixLoopFitFigure} is based on \eqn{SimplePrediction2} with the parameters
\begin{equation}
a_1 = 3.99 \,, \hskip 1.5 cm  b_1 =  1.74 \times 10^{-5} \,, \hskip 1.5 cm 
   c_1 =  9.77 \,.
\label{SixLoopFitParameters2}
\end{equation}
Interestingly, a nearly equally good fit is given by the following analytic
form, which contains remarkably simple constants,
\begin{equation}
\rho_L \simeq (\pi^2)^{L + 4/L - e^{\pi/2}}\,.
\end{equation}
Since we do not know the precise functional form, for the purposes
of extrapolating to higher-loop orders, it is useful to 
compare this to a different Ansatz,
\begin{equation}
\rho_L \simeq a_2 + b_2 c_2^L \,, 
\label{SimplePrediction}
\end{equation}
where again $a_2$, $b_2$ and $c_2$ are parameters.  The dashed curve
in \fig{SixLoopFitFigure} corresponds to \eqn{SimplePrediction} with
the parameters
\begin{equation}
a_2 = 0.179 \,, \hskip 1.5 cm  b_2 = 4.52  \times 10^{-4} \,, \hskip 1.5 cm 
c_2 =  7.30 \,.
\label{SixLoopFitParameters}
\end{equation}

Before extrapolating to higher loops, an interesting exercise is to 
use \eqns{SimplePrediction2}{SimplePrediction} to 
see how well they predict (or rather postdict) the
obtained UV divergences for $L=5,6$.
Because the Ans\"atze involves three parameters, we need to use the values for 
$L=2,3,4$ to fix the function.
For the anstaz in \eqn{SimplePrediction2} we obtain 
\begin{equation}
a_1 = 4.06 \,,   \hskip1.5cm
b_1 = 1.34 \times 10^{-5}\,, \hskip 1.5 cm  c_1 = 10.2\,.
\end{equation}
Plugging this solution into \eqn{SimplePrediction} gives good
predictions for the $L=5,6$ cases,
\be
\rho_5 \simeq 9.91 \,,  \hskip1.5cm
\rho_6 \simeq 74.0 \,.
\ee
Similarly, for the Ansatz in \eqn{SimplePrediction}, we obtain
\begin{equation}
a_2 = 0.127 \,, \hskip 1.5 cm  b_2 = 6.22 \times 10^{-4} \,, \hskip 1.5 cm 
  c_2 =  6.92 \,,
\end{equation}
and
\be
\rho_5 \simeq 9.99 \,,  \hskip1.5cm
\rho_6 \simeq 68.4 \,.
\ee
Presumably, the surprisingly good agreement between these approximate values
and the calculated ones in \eqns{SixLoopValue}{LowerLoopValues} is
somewhat accidental.  Nevertheless it does illustrate the remarkably
good predictive power of this simple extrapolation.  

Using the fit parameters based on explicit results through six loops,
we can easily predict approximate
values for higher loops.  For example, through $L=9$
from the Ansatz (\ref{SimplePrediction2}) we have,
\be
\rho_7 \simeq 540 \,, \hskip1.5cm
\rho_8 \simeq 4490 \,, \hskip1.5cm
\rho_9 \simeq 38700 \,,
\label{SimplePred2Values}
\ee
while the Ansatz  (\ref{SimplePrediction}) gives
\be
\rho_7 \simeq 500 \,, \hskip1.5cm
\rho_8 \simeq 3650 \,, \hskip1.5cm
\rho_9 \simeq 26700 \,.
\label{SimplePredValues}
\ee
The small numerical integration uncertainty from our $L=6$ result
feeds into this fit, propagating a few percent spread in the
estimates. Of course, the functional forms may be too naive, but
the two different fits gives an indication of the spread in predictions for
such extrapolations.

Another interesting numerical observation from
refs.~\cite{CK4l,FiveLoopNew} is that for an $SU(N_c)$ gauge group,
the ratio of the $1/N_c^2$-suppressed subleading-color contributions
to the leading color ones are fairly constant for $L=3,4,5$ and takes
a value of about $45$.  This observation immediately gives us a
prediction for the value of the divergence for the fully color-dressed
amplitude (including nonplanar contributions),
\begin{eqnarray}
{\cal A}_4^{(6)}\Bigr|_{\rm div} \hskip -.1 cm &\simeq&
 \frac{1}{\e} \, \frac{1}{(4\pi)^{15}} \, g^{14} \, s t A_4^\tree
  \, N_c^4 \, ( 68.68 \, N_c^2 + 3100 )
 \\
&& \hskip 1 cm \null \times
\Bigl[ s (\Tr_{1324} + \Tr_{1423}) + t (\Tr_{1243} + \Tr_{1342}) 
  + u (\Tr_{1234} + \Tr_{1432})  \Bigr] \,, \nn
\label{FinalVacuum}
\end{eqnarray}
where $\Tr_{1234} \equiv \Tr[T^{a_1} T^{a_2} T^{a_3} T^{a_4}]$ and we
assume that only the leading-color and $1/N_c^2$-suppressed single-trace
terms contribute, as is the case for $L=3,4,5$.

It is interesting to note that, at least through four loops, the
divergences of $\NeqEight$ supergravity are in the same critical
dimension $D_c=4+6/L$, and they are proportional to the same linear
combination of vacuum integrals as the $1/N_c^2$-suppressed terms of
MSYM~\cite{CK4l,FivePointBCJ}.  If this link between gravity and gauge
theories were to persist to all loop orders, then the
critical-dimension agreement alone would imply the four-dimensional
ultraviolet finiteness of the theory.  It would be very interesting to
directly check these ultraviolet divergence patterns in both gauge and
gravity theories at as high a loop order as possible, in order to see
if they could give insight into the UV properties of $\NeqEight$
supergravity, and into the precise values of the ultraviolet
divergences in MSYM in $D=4+6/L$, to {\em all} loop orders.


\section{Conclusions and outlook}
\label{ConclusionsSection}

We have shown that planar MSYM diverges in $D=5$ at six loops, in
accordance with expectations from previous explicit computations and
supersymmetry arguments.  This result raises various questions about
its relation to $(2,0)$ theory as discussed in
ref.~\cite{DouglasConjecture}. Because (2,0) theory is superconformal,
the $D=5$ MSYM UV cutoff must be related to the gauge coupling, and
combining this relation with the $S$-duality of the theory compactified
to $D=4$ should lead to strong constraints.  
Probably the simplest next step is to work out the $S$-dual
extension of the $D=5$ two-loop amplitude.

Through five loops, the planar MSYM four-point integrand has a 
manifestly nonzero behavior in the terms that control the UV divergence in the
expected critical dimension $D_c=4+6/L$.  Thus it is clear, without performing
any loop integrals, that the amplitudes diverge in the critical dimension.
At six loops, we were unable to find an integral representation 
in which all contributions are of the same sign.
Therefore we had to explicitly evaluate nontrivial integrals in order to
answer the question of whether MSYM diverges in $D=5$.  At six loops,
practical analytic techniques are not available for generic integrals,
so we resorted to numerical methods.  In particular, we used the sector
decomposition method as implemented in a modified version of the
{\sc FIESTA} program.

Our results show that, at least through six loops, the values of the
ultraviolet divergences in the critical dimension of the planar
amplitude approximately follow a simple exponential Ansatz.  Indeed,
extrapolating the results from two, three and four loops using this
Ansatz matches the calculated values at five and six loops remarkably
well.  The fact that our calculated six-loop value closely matches
this extrapolation gives us additional confidence that we have
computed the six-loop divergence correctly.  It also allows us to
extrapolate the value of the divergences in the critical dimension to
even higher loops.  Moreover, as also noted in
refs.~\cite{CompleteFourLoopYM,FiveLoopNew}, the ratios of the
numerical values of the $1/N_c^2$-suppressed terms to the
leading-color terms are approximately constant for $L=3,4,5$.
Assuming that this approximate constancy holds as well for $L=6$ gives
us a definite prediction for the value of the subleading-color
contributions to the divergence in $D=5$. While the origin of the
exponential behavior is still unclear, it does suggest that it might
be possible to understand the ultraviolet divergences of MSYM in the
critical dimension to {\em all} loop orders.

The same integration techniques described in this paper may be helpful
for resolving other problems.  An outstanding question that could be
resolved by computation of a higher-loop divergence is whether
$\NeqEight$ and other supergravity theories might be perturbatively
ultraviolet finite. (For recent reviews see
refs.~\cite{GravityUVReview}.) The current consensus for $\NeqEight$
supergravity is that a $D=4$ potential counterterm valid under all
known symmetries exists at seven
loops~\cite{SevenLoopGravity,VanishingVolume}. (A recent optimistic
opinion for all orders finiteness may be found in ref.~\cite{Kallosh},
while a pessimistic one may be found in ref.~\cite{Banks}.)  The same
potential counterterm could be studied in $D=24/5$ at five loops, which 
should be well within reach of the types of integration techniques
described here, once the supergravity integrand is constructed.  

Intriguingly, through at least four loops, the explicit values of the
$\NeqEight$ supergravity divergences (in the same critical dimension
as MSYM) are proportional to the same linear combination of
vacuum integrals as enter the subleading-color divergence of
$\NeqFour$ theory at the same loop order.  Moreover,
half-maximal supergravity appears to be better
behaved at three loops~\cite{ThreeLoopN4Grav} than had been
anticipated~\cite{VanishingVolume}.  These results emphasize the need
for further explicit computations at high loop orders, in order to help
unravel the ultraviolet properties of  supergravity theories.  
Our results demonstrate that evaluations of ultraviolet divergences are
feasible through at least six loops. Due to the close relation between
gravity and gauge-theory loop amplitudes~\cite{BCJLoop}, the results
presented here also provide a concrete initial step towards determining
the critical dimension of $\NeqEight$ supergravity at six loops.

As yet, there are no explicit forms of the $\NeqEight$ supergravity
integrands beyond four loops, although recent progress in the nonplanar
sector of MSYM at five loops~\cite{FiveLoopNew} 
suggests that the four-point five-loop amplitude of $\NeqEight$
supergravity is within reach.
In any case, our success at six loops with
the sector decomposition method suggests that although difficult,
an evaluation of the integrals likely to occur at seven loops could
be feasible.
How hard would a numerical evaluation at seven loops be, along the
lines discussed here?  As we discussed, a reasonable guess for the
number of sectors of a seven-loop integral is $1.6\times 10^8$.  If an
integral takes 2 seconds, then a 1000-core cluster can evaluate these
integrals in about 3 days.  This is perhaps a bit slow as we might
have hundreds of graphs and a more complex integrand, but with further
optimization and a larger cluster even this computation should come
within reach.  It may also be possible to achieve further gains based on
converting the vacuum integrals to propagator integrals and
factorizing them into products of lower-loop
integrals~\cite{Vladimirov}, as has been applied recently in maximally
supersymmetric theories at four and five
loops~\cite{CK4l,FiveLoopNew}.  There are also other methods for
attacking this problem, such as the powerful DRA method~\cite{DRA},
which offers much higher precision than can be obtained by sector
decomposition, provided that an appropriate large system of linear
equations can be solved symbolically.

In summary, in this paper we showed that maximally supersymmetric
Yang-Mills theory diverges at six loops in $D=5$, settling the
question of whether the link to the $(2,0)$ theory might imply an
improved UV behavior.  We showed that, even at loop orders as high as
six, and possibly higher, we can directly determine the UV properties
of supersymmetric gauge and gravity theories.  We also uncovered a
simple approximate exponential pattern for the values of the
divergences in the critical dimension where they first occur. 
This pattern may provide clues toward unraveling the all-loop-order
UV structure.  It is not obvious how to reconcile the appearance of a
six-loop divergence in $D=5$ MSYM with the finiteness of its UV
completion, the (2,0) theory in $D=6$.  Presumably, the divergence must
be cutoff by additional degrees of freedom in the UV theory.  As
discussed in refs. \cite{DouglasConjecture,Lambert}, there are already
candidates for these degrees of freedom as nonperturbative states in
the $D=5$ theory, so that it may be possible to understand this in
$D=5$ terms.  Perhaps the simplest conjecture is that $S$-dual
extensions of the $D=5$ amplitudes (as discussed in
ref.~\cite{DouglasConjecture}) are finite to all orders.  

\section*{Acknowledgments}
We thank Nima Arkani-Hamed, Radu Roiban and Edward Witten for helpful
discussions.  We also thank Paul Heslop and Gregory Korchemsky for
assistance in comparing our amplitude integrand to their
form~\cite{EdenRep}.  We thank the Institute for Nuclear Theory in
Seattle and the Banff International Research Station for hospitality,
and Academic Technology Services at UCLA for computer support.  This
research was supported by the US Department of Energy under contracts
DE--AC02--76SF00515 and DE--FG03--91ER40662.  JJMC acknowledges that
this publication was made possible through the support of the Stanford
Institute for Theoretical Physics, and a grant from the John Templeton
Foundation. The opinions expressed in this publication are those of
the authors and do not necessarily reflect the views of the John
Templeton Foundation.


\end{document}